\documentstyle[prb,aps]{revtex}
\begin{document}
\preprint{UCF-CM-96-004}
%
%
\title{Spin Ensemble Density Functional Theory for Inhomogeneous Quantum Hall 
Systems.}
\author{M.I. Lubin, O.Heinonen, and M.D. Johnson }
\address{Department of Physics, University of Central Florida, Orlando, 
Florida 32816-2385}
\maketitle
\begin{abstract}
We have developed an ensemble density functional theory which includes 
spin degrees of freedom for nonuniform quantum Hall systems.  We have applied this 
theory using a local-spin-density approximation to study the 
edge reconstruction of parabolically 
confined quantum dots.  
For a Zeeman splitting below a certain critical value,  the edge of 
completely polarized maximum density droplet  reconstructs into a spin-unpolarized 
structure.  For larger Zeeman splittings, the edge remains polarized and 
develops an exchange hole. 
\end{abstract}
\section{Introduction}
\label{intro}

The recent development of nanofabrication technology has made it possible to 
manufacture semiconductor systems with reduced dimensionality and very high   
electron mobility. These technological advances have led to the discovery of such 
fascinating phenomena as the fractional and integer quantum Hall effects 
(QHE)~\cite{vonKlitzing}.  These occur in a two-dimensional electron gas in a  
magnetic field ${\bf B}=B \hat{z}$ perpendicular to the electron system~\cite{QHE}.  
A quantum treatment of the motion of an infinite, homogeneous system shows that 
the kinetic energy takes discrete values $(n+1/2)\hbar\omega_c$,  where $n$ is the  
Landau level  index ($n=0,1,2,\ldots$) 
and $\omega_c=eB/m^{\star}c$ is the cyclotron frequency.  Each 
Landau level contains $n_B=B/\Phi_0$ states per unit area, or one state for each 
magnetic flux quantum $\Phi_0=h c/e$, giving rise to a macroscopic  Landau level 
degeneracy. The ratio of the electron areal density  $n({\bf r})$ to $n_B$ defines the 
filling factor $\nu({\bf r}) = n({\bf r})/n_B$. The filling factor  also can be 
expressed as $\nu = 2\pi l_B^2 n$, where $l_B=\sqrt{\hbar c /eB}$ is the magnetic 
length.

The fractional quantum Hall effect can occur when electron-electron 
interactions dominate disorder.  At certain filling factors of the form $\nu = p/q$, 
with $p$ and $q$ relative primes and $q$ odd, electron-electron interactions cause 
the condensation of the electrons into highly correlated states. These states are 
incompressible with an energy gap separating the ground state from the bulk excited 
states.  However, in a finite system, there must exist  gapless excitations 
localized near the edges~\cite{Halperin_gapless}.  Thus, the low-energy physics of 
finite systems is dominated by the gapless edge modes.  Therefore, in is necessary
to be able to accurately model edges of FQHE systems in order to explain
experiments. In a  finite FQHE system 
with 
the potential confining the electrons varying slowly compared to $l_B$, the 
electronic structure at the edges may form a series of alternating compressible and 
incompressible regions with a step-like density profile~\cite{Beenakker,Chklovskii}.
In addition to standard transport measurements, there are now a variety
of probes to directly study edge structures in inhomogeneous systems.  
Examples are capacitance spectroscopy of the quantum Hall edges~\cite{Zhitenev1}, 
time-resolved measurements of edge magnetoplasmons\cite{Zhitenev2},
and surface acoustic waves techniques which are capable of 
resolving very small spatial inhomogeneities in the electron density~\cite{Shilton}. 
  Addition spectroscopy has also been used to study quantum dots  with
sizes of the order of 100 nm and with 10 to 100 electrons~\cite{McEuen}.
   
For an explanation of experimental studies it is highly desirable to have a 
computational approach which accurately treats systems with of the order of 
$1 - 10^3$ electrons, and which can include  effects such as
accurate confinements, spin degrees of freedom, and finite layer thickness.
  Exact numerical diagonalizations are limited to very small systems
 ($ N \leq 10$~)~\cite{Yang,Ahn}.  Semiclassical  
 methods\cite{Beenakker,Chklovskii} do not 
accurately treat electron-electron interactions, and  effective field 
theories~\cite{Fradkin}  cannot give accurate quantitative information about
many system properties.
 A method that can deal with larger number of electrons is the composite fermion   
theory in the Hartree approximation~\cite{Goldman}. However,  in this approach, the 
singular Chern-Simons gauge field is replaced by its smooth spatial average, and the 
composite fermion mass has to be put into the calculations by hand.  
Furthermore, interpretation of the results is sometimes difficult and
ambiguous.  On the other hand, density functional theory (DFT) is known as a  
general quantitative method to include exchange-correlation effects in inhomogeneous 
systems without any fitting parameters. In this paper we show that it can be used to 
give highly accurate results for quantum Hall systems.
Preliminary results for spin polarized systems were reported
earlier~\cite{Heinonen1,Heinonen2}.
 
  The DFT was originally formulated by Hohenberg and Kohn
 as a practical method for a description of the ground state properties
of many-body systems~\cite{Hohenberg} .  The foundation of DFT is the
Hohenberg-Kohn theorem, which states that the ground state density uniquely
determines the Hamiltonian of a system.  Furthermore, a variational principle
states that the ground state density minimizes the energy of the system.  
We will use the  constrained search  formulation of Levy~\cite{Levy0} for the 
Hohenberg-Kohn theorem and its associated variational principle.
In this elegant approach the ground state energy $E$ can be written as a functional 
of density 
\begin{equation}
E[n] = F[n] + \int d {\bf r}\, n({\bf r}) V_{\rm ext} ({\bf r}).
\label{HK1}
\end{equation}
 Here
\begin{equation} 
F[n] =\inf_{\Psi\rightarrow n}\langle \Psi|\hat{T} + \hat{V}_{\rm ee}|\Psi\rangle,
\label{HK2}
\end{equation}
with $\hat{T}$, $\hat{V}_{\rm ee}$, and $\hat V_{\rm ext}$ kinetic energy,  
electron-electron interactions, and external potential, respectively. The infimum is 
taken over all many-body states $\Psi$ that
yield a fixed density $n({\bf r})$. $F[n]$ so defined is then a universal 
functional of the density 
$n({\bf r})$.  For a given external potential $V_{\rm ext}$, the true ground state 
density is the function $n({\bf r})$ which minimizes $E[n]$ in Eq. (\ref{HK1}).

The origins of the DFT are to be found in the statistical method developed by 
Thomas and Fermi~\cite{Thomas}.  
They first realized the advantage of describing an inhomogeneous systems by using 
the density 
\begin{equation}
  n({\bf r})=\int d {\bf r}_2\ldots\int d {\bf r}_N
\mid\Psi({\bf r},{\bf r}_2,\ldots,{\bf r}_N)\mid^2,
\label{wavefunction}
\end{equation}
an observable, rather than the  unobservable complex wave function $\Psi$ of $Nd$
variables in $d$ dimensions. Thomas-Fermi theory is a way 
to find an approximate $n({\bf r})$.
The theory is valid in the semi-classical limit  and  has a successful history of 
applications to many different problems~\cite{March}.
But since Thomas-Fermi method  neglects exchange-correlation effects and makes an 
approximation for the kinetic energy functional, it has some serious deficiences as 
well.  For example, Thomas-Fermi theory cannot predict ferromagnetism.
DFT remedies these problems by explicitly (and formally exactly) incorporating 
exchange-correlation effects as well as interaction parts of the kinetic energy 
functional into an exchange and correlation energy functional $E_{\rm xc}[n({\bf r})]$, 
and by developing a useful computational scheme for including exchange-correlation 
effects~\cite{Kohn}.  This is done by introducing
an auxiliary non-interacting system with a ground-state density $n_s({\bf r})$, 
and by asserting that there exists an effective potential $V_{s}({\bf r})$ for this 
system such that $n_s({\bf r})=n({\bf r})$, with $n({\bf r})$ the ground-state 
density of the real, interacting system. A system with this property
is called $v$-representable. The density is then obtained from a simple 
Slater-determinant of the so-called Kohn-Sham (KS) orbitals 
$\psi_{\alpha}({\bf r})$, 
$n_s({\bf r})=\sum_{\alpha=1}^N|\psi_\alpha({\bf r})|^2$, 
where $\psi_\alpha({\bf r})$ are obtained by self-consistently solving the KS 
equations~\cite{Kohn}
\begin{equation}
{h}_{\rm eff}\psi_\alpha({\bf r})=
\left[T + V_s({\bf r})\right]\psi_\alpha({\bf r})
 =\varepsilon_\alpha\psi_{\alpha}({\bf r}).
\label{Kohn-Sham}
\end{equation}
The self-consistency is achieved in practice by iteratively obtaining the 
eigenstates and occupying the $N$ eigenstates with the lowest eigenvalues
$\varepsilon_\alpha$.
 The effective potential ${V}_s({\bf r})$  can be derived from 
the  Hohenberg-Kohn theorem and its associated variational principle. First, 
the functional $E[n]$ [Eq.~(\ref{HK1})] can be decomposed as
\begin{equation}
E[n]=T_0[n] + \int d {\bf r} n({\bf r}) \left[V_{\rm ext}({\bf r})
+ \frac{1}{2}V_{\rm H}({\bf r})\right] + E_{\rm xc}[n].
\label{decomposition}
\end{equation}
Here $T_0$ is the kinetic energy of a non-interacting system with density $n$, 
$V_{\rm H}$ is 
the classical (Hartree) Coulomb potential
\begin{equation}
V_{\rm H}({\bf r})=\frac{e^2}{\epsilon_0}\int d {\bf r'}\frac{n({\bf r'})}
{|{\bf r}-{\bf r'}|},
\end{equation}
with $\epsilon_0$ the static dielectric constant,
and $E_{\rm xc}$ may be viewed as a definition of the exchange-correlation energy.  
The  non-interacting kinetic energy $T_0$ is treated exactly in this approach,  
which removes many of the deficiencies of the Thomas-Fermi model.  The variational 
principle applied to Eq.~(\ref{decomposition}) yields
\begin{equation}
\frac{\delta E[n]}{\delta n({\bf r})} =
 \frac{\delta T_0}{\delta n({\bf r})}
+ V_{\rm ext}({\bf r}) + V_{\rm H}({\bf r})+V_{\rm xc}({\bf r})=\mu,
\label{variational}
\end{equation}
  where $\mu$  is the Lagrange multiplier associated with the requirement
of constant particle number, and the exchange-correlation potential $V_{\rm xc}$ is 
formally defined as the functional derivative
\begin{equation}
V_{\rm xc}({\bf r}) \equiv \frac{\delta E_{\rm xc}[n]}{\delta n({\bf r})}.
\label{xc_potential}
\end{equation}
Comparison of Eq. (\ref{variational}) with the corresponding relationship for a 
non-interacting system, 
\begin{equation}
\frac{\delta E[n]}{\delta n({\bf r})} =
 \frac{\delta T_0}{\delta n({\bf r})}
+ V_s({\bf r}) =\mu ,
\end{equation}
gives an expression for the effective potential
\begin{equation}
V_s({\bf r}) = V_{\rm ext}({\bf r}) +V_{\rm H}({\bf r})+ V_{\rm xc}({\bf r}).
\label{eff_potential}
\end{equation}
In Eq. (\ref{decomposition}) all terms but the 
exchange-correlation energy $E_{\rm xc}$ 
can be evaluated exactly.
In practical calculations, the local density approximation (LDA) is often 
used~\cite{Kohn}. In this approximation, one writes
\begin{equation}
E_{\rm xc}^{\rm LDA}=\int d {\bf r}\, n({\bf r})\epsilon_{\rm xc}(n({\bf r})),
\label{energy_LDA}
\end{equation}
where $\epsilon_{\rm xc}(n)$ is the exchange-correlation energy per particle of an 
infinite, homogeneous system of density $n$. 
The exchange-correlation potential
[Eq. (\ref{xc_potential})] is then obtained as 
\begin{equation}
V_{\rm xc}^{\rm LDA}({\bf r}) = \left.\frac
{d\left[n\epsilon_{\rm xc}(n)\right]}
{d n}
\right|_{n =n(r)}.
\label{xc_potential_LDA}
\end{equation}

Above we ignored the electron spin -- $n({\bf r})$ was the total electron density, 
and the spin degree of freedom was neglected. The Hohenberg-Kohn
theorem formally ensures that {\em every} property, including the
spin density or polarization, can be obtained
from the ground-state density. However,  practical LDA calculations of systems with 
spontaneously broken symmetries, such as spin rotation symmetry, typically are
much improved if the order parameter of the broken symmetry, {\em e.g.,}
spin density or polarization, is explicitly included by construction. In particular,
the broken symmetry may not otherwise be obtained accurately from the LDA.
In GaAs samples, where most of QHE experiments have been done, the spin degree
of freedom is important, and may lead to inhomogeneous spin densities.
This is because the effective  Zeeman 
energy  $g^*\mu_B B$ is quite small compared to the cyclotron energy 
$\hbar\omega_c$,
$g^*\mu_B B/\hbar\omega_c \approx 0.02$, 
due to the small effective mass $m^*=0.068 m_e$ and a reduced Land\'{e} g-factor  
$|g^*| = 0.44$. This means that in a uniform noninteracting system, two highly 
degenerate Zeeman levels with the same Landau level index $n$ are almost degenerate 
in energy.  
Since the dielectric constant for GaAs $\epsilon_0\approx 13$, the cyclotron energy 
$\hbar\omega_c$ and Coulomb energy $e^2/(\epsilon_0 l_B)$ are of
the same order of magnitude for magnetic fields of the strength of a few Tesla  
($\frac{\hbar\omega_c}{e^2/(\epsilon_0 l_B)}\approx 0.4\sqrt{B[{\rm T}]}$).
Using the Coulomb energy as the unit of energy, 
the dimensionless parameter characterizing Zeeman coupling can 
then be defined as
\begin{equation}
\tilde g \equiv \frac{g^*\mu_B B}{e^2/(\epsilon_0 l_B)}.
\label{g_def}
\end{equation}
The small value of $\tilde g$ (typically about 0.02) makes the existence of partly 
polarized states energetically possible~\cite{Halperin,Zhang} even at $\nu < 1$.
Therefore, it is necessary to include the spin degree of freedom in any
quantitative theoretical approach to QHE systems.  A reasonable first step is to 
generalize the DFT to include the spin polarization\cite{Barth,Gunnarsson}.  
An exact treatment of the spins, in general, requires~\cite{Barth} 
the replacement of the charge 
density $n({\bf r})$  by the single-particle density matrix 
$
\rho_{\sigma\sigma'}({\bf r})=
    \langle 
       0|\hat\psi^{+}_{\sigma}({\bf r})\hat\psi_{\sigma'}({\bf r})|0
    \rangle
$.
Here, $\hat\psi^{+}_{\sigma}({\bf r})$ and
$\hat\psi_{\sigma}({\bf r})$ are the usual field operators corresponding
to the annihilation and creation of an electron
with spin $\sigma$ at ${\bf r}$, and $|0\rangle$ is the ground state of the system. 
With a constant magnetic field applied in the $z$-direction, the $\hat z$ 
component of the total spin angular momentum is a constant of the motion and it
is convenient to assume that the magnetization density only has a $\hat z$
component. Under this assumption,
the single-particle density matrix can be taken 
to be diagonal, 
$\rho_{\sigma\sigma'}({\bf r})=\rho_{\sigma\sigma'}({\bf r})\delta_{\sigma\sigma'}$.
In this case the constrained search procedure of Eqs.~(\ref{HK1}) and (\ref{HK2})
is modified to~\cite{Perdew}
\begin{equation}
E[n_{\uparrow},n_{\downarrow}] = F[n_{\uparrow},n_{\downarrow}] + 
\int d {\bf r}\, n({\bf r}) V_{\rm ext} ({\bf r}),
\label{HKS1}
\end{equation}
 where
$ F[n_{\uparrow},n_{\downarrow}] =
\inf_{\Psi\rightarrow n}
  \langle \Psi|\hat{T} + \hat E_{\rm Z}+ \hat{V}_{\rm ee}|\Psi
  \rangle$ and $\hat E_{\rm Z}$ is the Zeeman energy with $\Psi$ 
  yielding fixed densities 
$n_{\sigma}({\bf r})$ .
The local spin-density approximation (LSDA) is then given by
\begin{equation}
E_{\rm xc}^{\rm LSDA}[n_{\uparrow},n_{\downarrow}]=
   \int d {\bf r}\, n({\bf r})  
    \epsilon_{\rm xc}[n_{\uparrow}({\bf r}),n_{\downarrow}({\bf r})],
\label{LSDA}
\end{equation}
where $\epsilon_{\rm xc}[n_{\uparrow},n_{\downarrow}]$ is the exchange-correlation 
energy per particle in a homogeneous system with up- and down- spin densities 
$n_{\uparrow}$ and $n_{\downarrow}$, respectively.
In spite of the LSDA being justified only in the limit of small spatial variations 
of the electron density, this approximation has been surprisingly successful in 
describing the properties of inhomogeneous atomic, molecular and solid-state 
systems~\cite{Jones}.  
This scheme correctly predicted, for example, ferromagnetism in Fe, Co and Ni among 
the transition metals~\cite{Jones}.  Moreover, the self-interaction-corrected LSDA 
was successfully applied to some strongly correlated systems such as the 
transition-metal oxides and a Hubbard model representing a ${\rm CuO_2}$ layer in 
the cuprate superconductors~\cite{Svane}.

In the original Kohn-Sham formulation [Eqs.~(\ref{HK1}) and (\ref{HK2})], the ground 
state was assumed to be nondegenerate and
the ground state density $n({\bf r})$ assumed to be pure-state  
$v$-representable. This means that $n({\bf r})$ can be be expressed in terms of a 
single Slater determinant of KS orbitals $\psi_{\alpha}({\bf r})$ obeying an 
effective single-particle Schr\"odinger equation~[Eq.~(\ref{Kohn-Sham})].
However, there are systems which are known not to be $v$-representable in
this sense. One class of such systems was considered independently by
Levy~\cite{Levy} and Lieb~\cite{Lieb}.  Consider a system with $q$  independent 
$N$-particle degenerate ground states
$
|\Psi_1\rangle,\ldots,|\Psi_q\rangle.  
$
  Then construct the density matrix
\begin{equation}
\hat{D} = \sum^q_{i=1} d_i |\Psi_i\rangle\langle\Psi_i|
\label{density_matrix}
\end{equation}
 with $d_i=d_i^*\geq 0$, $\sum^q_{i=1}d_i=1$. This yields the density
\begin{equation}
n({\bf r}) = {\rm Tr}\{ \hat{D} \hat{n}({\bf r}) \}=\sum^q_{i=1}d_i n_{i}({\bf r}),
\label{expansion}
\end{equation}
 where $n_{i}({\bf r}) = \langle\Psi_i|\hat{n}({\bf r})|\Psi_i\rangle$ 
(spin degrees of freedom are neglected for simplicity).
As was proven by Levy and Lieb, if  $q > 2$ the density $n({\bf r})$ cannot be 
represented by a single ground state in DFT, {\em i.e.,} it cannot be obtained by a single 
Slater determinant of the $N$ lowest energy KS orbitals. 
  However, there exists a generalization of Hohenberg-Kohn theorem which
provides a one-to-one correspondence between a ground state density
$n({\bf r})$ and the Hamiltonian even for system with a ground state density
which can be of the form of Eq.~(\ref{expansion}).
By extending the functional $F[n]$ in Eq.~(\ref{HK2}) to
\begin{equation} 
F_{\rm E}[n] =\inf_{\hat{D}\rightarrow n}{\rm Tr}
\{\hat{D}(\hat{T}+\hat{V_{\rm ee}})\},
\label{HKE}
\end{equation}
with the infimum taken over all $\hat{D}$ yielding a fixed density $n({\bf r})$,
 there is then a generalized variational principle which states that
$F_{\rm E}[n]$  is minimized by the ground state density, which can now be represented
by an ensemble of wave functions, even  if it cannot be represented by a
single Slater determinant.  This generalization is called 
 {\it ensemble} density functional theory.  As we shall see below, fractional
QHE systems are not $v$-representable, so an ensemble DFT has to be used.
 
In Sec.~\ref{ensemble} we review our ensemble DFT scheme. 
The essential features of our ensemble DFT approach will be illustrated in 
Sec.~\ref{spdft} by applying it to spin-polarized quantum Hall dot.
Sec.~\ref{sudft} then describes the spin-unpolarized edge reconstruction of $\nu=1$ 
quantum Hall dot. The phase diagram for the spin textured edge reconstruction of the 
maximum density droplet  will be presented there.
Finally, conclusions are given in Sec.~\ref{concl}.

\section{Practical algorithm for ensemble density functional theory}
\label{ensemble}

 In practical ensemble DFT calculations one introduces as in the KS scheme an 
auxiliary non-interacting system which provides the basis for the density
matrix and has a ground state density identical to the 
interacting system at hand.  By using the variational principle, one then arrives at 
a set of equations analogous\cite{Dreizler} to
the KS equations Eq.~(\ref{Kohn-Sham}).
  However, the density for $N$ electrons is now given  by
\begin{equation}
n({\bf r})=\sum_{\alpha}f_{\alpha}|\psi_{\alpha}({\bf r})|^2,
\quad \sum_{\alpha}f_{\alpha}=N,
\label{expansion2}
\end{equation}
with the occupation numbers $f_{\alpha}$  in the interval $0\leq f_{\alpha}\leq 1$.
One obtains fractional occupancies $f_{\alpha}$ only when the corresponding KS 
eigenvalues
$\varepsilon_{\alpha}$ are degenerate and equal to the Fermy energy $\varepsilon_F$.
(If $\varepsilon_\alpha<\varepsilon_F$, then $f_\alpha=1$.) 
Let us show briefly why applying DFT to the FQHE inevitably requires ensemble DFT.  
We consider an infinite, homogeneous fractional QHE system at a filling factor
of $\nu=1/3$, and assume that we have the {\em exact} exchange-correlation
potential $V_{\rm xc}$ for this system. We then construct 
a set of determinantal $N$-particle
wavefunctions $\left\{\Psi_i\right\}$ made up from the KS orbitals 
$\psi_{\alpha}$ obtained
by solving the KS equation using the exact exchange-correlation potential.
Label each member $\Psi_i$ of this set 
by the set of numbers $\{\theta_{i\alpha}\}$ identifying the
KS orbitals used to construct $\Psi_i$ [$\theta_{i\alpha}=(0)1$ for 
(un)occupied orbitals, and $\sum_{\alpha}\theta_{i\alpha}=N$].  Because the
KS orbitals are orthogonal, a state $\Psi_i$ has density~\cite{Mahan}
\begin{equation}
n_{i}({\bf r}) =\sum_{\alpha}\theta_{i\alpha}|\psi_{\alpha}({\bf r})|^2.
\label{expansion1}
\end{equation}
Construct three states $\Psi_i$, $i=1,2,3$, with occupancies 
$\{\theta_{i\alpha}\}=\{100100\ldots\}$, $\{010010\ldots\}$, $\{001001\ldots\}$.
The occupied KS orbitals are all degenerate lowest Landau level
states (since the system is homogeneous), 
so the three $\Psi_i$ are degenerate.  Then construct a density matrix 
of the form of  
Eq.~(\ref{density_matrix}) with $d_1=d_2=d_3=1/3$.  The resulting density
$n({\bf r})$ (Eq.~\ref{expansion}) is constant: 
$n({\bf r})=\frac{1}{3}\frac{1}{2\pi l_B^2}$, corresponding to a uniform
system with $\nu=1/3$.  Now we can appeal to the Levy-Lieb
theorem~\cite{Levy,Lieb}:
since the $\nu=1/3$ ground state density can be constructed as in
Eqs. (\ref{density_matrix}) and (\ref{expansion}) with $q>2$ degenerate states
$\Psi_i$ this density is therefore {\it not} pure state $v$-representable, and 
ensemble DFT must be used.

In inhomogeneous FQHE systems, not all KS orbitals are degenerate, but some
are.  By a simple extension of the argument in the paragraph above, such systems too 
are not  pure state $v$-representable, and ensemble DFT must be used.
For inhomogeneous systems one finds $M$ orbitals with 
$\varepsilon_{\alpha}< \varepsilon_F$ and $D$ degenerate orbitals with
$\varepsilon_{\alpha}= \varepsilon_F$.  One constructs determinantal 
wavefunctions $\Psi_i$ in which all $M$ low-energy orbitals are occupied;
the $\Psi_i$ differ by which $N-M$ of the $D$ degenerate orbitals are occupied.
Using Eq. (\ref{expansion}) and the density of determinantal wavefunctions given
in Eq.~(\ref{expansion1}), the total density for the ensemble represented by
$\hat{D}$ can be calculated as
\begin{equation}
n({\bf r}) =\sum_{\alpha}\sum^q_{i=1}d_i \theta_{i\alpha}|\psi_{\alpha}({\bf r})|^2.
\label{expansion3}
\end{equation}
Comparing  the result with Eq.(\ref{expansion2}), one can see how the fractional 
occupational numbers $f_{\alpha}$ of the degenerate KS orbitals follow from the 
weights $d_i$ in the expansion of density matrix $\hat{D}$:
\begin{equation}
f_{\alpha} = \sum_{i=1}^q d_i \theta_{i\alpha}.
\label{weights}
\end{equation}
As we mentioned in Sec.~\ref{intro}, for the density defined by 
Eq.~(\ref{expansion3}),
a generalization of Hohenberg-Kohn theorem exists  and an extended variational 
principle [Eq.~(\ref{HKE})] can be used.  However, a procedure to compute the 
fractional occupancies $f_{\alpha}$ has not existed~\cite{Gross}, and one major 
advance in our  work is that we have found a simple way to generate the occupancies, 
at least for the FQHE~\cite{Heinonen1,Heinonen2}.  
Applying ensemble DFT to the FQHE, we have 
found that fractionally-occupied KS orbitals are indeed degenerate at the Fermi 
energy, consistent with our demonstration above that the FQHE is in general not 
pure-state $v$-representable.
We will review this scheme here.

In our algorithm, we start with a set of input occupancies and single-particle 
orbitals and
iterate the system $N_{\rm eq}$ times using the KS scheme. 
The number $N_{\rm eq}$ is chosen large enough (about 20-40 in practical 
calculations) that the density is close to the final density after $N_{\rm eq}$ 
iterations. If the density of the system could be represented by a single
Slater determinant of the KS orbitals, we would now essentially
be done.
However, in an ``ensemble  $v$-representable'' (but not pure state 
$v$-representable) system there are  many degenerate or near-degenerate KS orbitals 
at the Fermi energy, and small fluctuations in the density between iterations cause 
the KS scheme to occupy a different subset of these orbitals each iteration. This 
corresponds to constructing different Slater determinants $\Psi_j$ each iteration.  
In other words, when the KS orbitals are degenerate at the Fermi energy there is an 
ambiguity in how to occupy these degenerate orbitals. Nevertheless, by  associating 
the number of the iteration $j$ with the particular Fock state  $\Psi_j$  
represented by the occupation numbers $\theta_{j\alpha}$, one can accumulate the 
running average  occupancies after each iteration 
\begin{equation} 
f_{\alpha, N_{\rm it}}=\frac{1}{N_{\rm it} - N_{\rm eq}}
\sum_{j=N_{\rm eq}+1}^{N_{\rm it}}\theta_{j\alpha},
\label{average3}
\end{equation}
where $N_{\rm it}$ is the number of iterations. 
This is clearly of the form of Eq.~(\ref{weights}) with the weights $d_i$ given
by the relative frequency of the corresponding Slater determinant during
$N_{\rm it} - N_{\rm eq}$ iterations.
To achieve a self-consistent solution, the system of KS equations is then iterated 
using average  occupancies $f_{\alpha, N_{\rm it}}$ instead of $\theta_{j\alpha}$, 
with the density after $N_{\rm it}$ iterations calculated as
\begin{equation}
n_{N_{\rm it}}({\bf r})=
\sum_{\alpha}f_{\alpha, N_{\rm it}}|\psi_{\alpha}({\bf r})|^2.
\end{equation}
This density is used to calculate the new effective potential ${V}_s({\bf r})$
according Eq.(\ref{eff_potential}), and by solving the resulting KS equations a  new 
set of occupancies $\theta_{j\alpha}$ is obtained, which can be again used to update 
averages $f_{\alpha, N_{it}}$. When a self-consistent solution is
obtained this highly nonlinear map should give the same input and output densities 
after one iteration.
To prove formally that this scheme with the running average  occupancies
converges self-consistently to the physical density is by no means a trivial 
problem.  However,
  we have numerically verified for FQHE systems that a finite-temperature version of 
our scheme converges to a thermal ensemble at finite temperatures down to 
temperatures of the order of $10^{-4}\hbar\omega_c/k_B$\cite{Heinonen2}. We
have also performed  Monte Carlo simulations about the ensemble obtained by our 
scheme. In these simulations, we used a Metropolis algorithm to randomly change the 
occupation numbers about our converged solution, keeping the chemical potential 
fixed. The free energy of the new set of occupation numbers was calculated 
self-consistently. The results were that to within numerical
accuracy our scheme gives the lowest free energy. We have also checked our
ensemble DFT against small system numerical diagonalizations with very
good results (see Sec. \ref{spdft}).

This algorithm has made ensemble DFT a practical calculation tool, and it may be 
possible to apply it to strongly correlated systems other than the QHE, for example 
inhomogeneous Mott insulators, provided there are accurate approximations for the
ground state energies of homogeneous systems available.

\section{Spin-polarized quantum dot in a FQHE regime}
\label{spdft}

In order to illustrate the essential features of our ensemble DFT scheme, we first 
neglect  the spin degree of freedom and consider a spin-polarized 
quantum dot in the FQHE regime. Typical quantum dots  contain about 10-100 
electrons, so the quantum dot can be used as a model system in order to
demonstrate the usefulness of the DFT in the study of large inhomogeneous electron 
systems.
 Moreover, quantum dots are believed to have highly correlated ground states  in 
strong magnetic fields~\cite{Maksym}.  Hence, these systems can be used to show how 
well these  strong correlation effects are represented by  LDA compared to exact 
diagonalization studies~\cite{Yang,Ahn}. There have been some attempts to use
density functional theory to model such systems. 
 Ferconi and Vignale~\cite{Ferconi1} performed current-spin DFT studies of quantum 
dots in the integer QHE regime with a small number of electrons. In their
calculations, the energy gaps due to correlation effects were not included,
and the spin-dependence of the exchange-correlation energy was taken to be
that of an electron gas in the $B\to0$ limit. Their results for a spin-polarized
three-electron system are in good agreement with results from
exact diagonalizations. However, their approach cannot be extended to
include fractionally occupied states or the complicated spin dependence
of the exchange-correlation energy in the strong magnetic field region. 

Electron-electron interactions in the ground state of quantum dots have been  
studied experimentally, for example by measuring the tunneling conductance through a 
Coulomb island~\cite{McEuen,Schmidt}.
The dots used experimentally can very accurately be modeled as 
parabolic~\cite{McEuen}, i.e., the Coulomb islands are confined by a parabolic  
potential $V_{ext} = m^*\Omega^2 r^2/2$, where $m^*$ is the effective mass of an 
electron, $\Omega$ characterizes the strength of confining potential and $r$ is the 
distance from the center of the dot.  We chose  this potential for our model system.

Due to the circular symmetry we can label the KS orbitals $\psi_{\alpha}({\bf r})$ 
by  Landau level index $n\geq0$, and by angular momentum label $m\geq -n$ ($\alpha 
\equiv \{m,n\}$) and expand $\psi_{\alpha}({\bf r})$ in the eigenstates $|mn\rangle$ 
of the single-particle  Hamiltonian  $\hat T_0=(-i\hbar\nabla + \frac{e}{c}{\bf 
A})^2/2m^*$ as 
\begin{equation}
\psi_{mn}({\bf r})=e^{im\phi}\varphi_{mn}(r)=
\sum_{n'} C_{mnn'}\langle {\bf r}|mn'\rangle.
\end{equation}
 We can then solve self-consistently the KS equations 
${h}_{\rm eff}\varphi_{mn}=\varepsilon_{mn}\varphi_{mn}$
for the radial parts $\varphi_{mn}$ of KS orbitals for each value of
the angular momentum $m$ separately. 
The orthonormal basis $\langle {\bf r}|mn\rangle$  can be written in the cylindrical 
gauge ${\bf A}=\frac{1}{2}B(x,-y,0)$ 
in terms of the associated Laguerre polynomials~\cite{Fock} $L^m_n$
\begin{equation}
\langle {\bf r}|mn\rangle = 
            \frac{1}{\sqrt{2\pi l_B^2}}
            \left(
               \frac{n!}{(n+m)!}
            \right)^{1/2}
           \left(
              \frac{r}{\sqrt{2} l_B}
           \right)^m
           L^m_n
           \left(
              \frac{r^2}{2 l_B^2}
           \right)
              e^{im\phi - r^2/4 l_B^2}.
\label{Laguerre}          
\end{equation}
These are  centered on circles of radii 
$r_{mn}\approx\sqrt{2(m+n)}\l_B$ with Gaussian fall-offs for $r\ll r_{mn}$ and $r\gg 
r_{mn}$. 
We write the density $n({\bf r})$  in dimensionless form as a nonuniform 
filling factor  $\nu({\bf r}) = 2\pi l_B^2 n({\bf r})$,
\begin{equation}
\nu=2\pi l_B^2 \sum_{mn} f_{mn} |\psi_{mn}|^2=
\sum_{mnn'n''}  f_{mn} C_{mnn'}C_{mnn''}\langle mn' |mn'' \rangle.
\end{equation}
This is used to calculate the effective potential $V_s(r)$ in the KS equations. In 
the lowest Landau level these expressions simplify (for example, the filling factor 
becomes just 
$\nu(r)=e^{-x}\sum_m f_{m0}\frac{x^m}{m!}$ with  $x=r^2/4l_B^2$). However,  
we have used the four lowest Landau levels ($n=0,\ldots,3$) and kept the
general expressions, since as we mentioned in Sec.~\ref{intro}, the cyclotron energy 
$\hbar\omega_c$ and Coulomb energy $e^2/(\epsilon_0 l_B)$ are of
the same order of magnitude for typical experimental situation, so there
can be appreciable Landau-level mixing.
 The solution of the KS equations in the basis $|mn\rangle$ reduces to iteratively 
obtaining the eigenstates  with the lowest energies of the following block-diagonal 
Hamiltonian matrix
\begin{equation}
h_{{\rm eff},mn'n} = \hbar\omega_c(n+\frac{1}{2})+ \langle mn'|\hat V_s|mn 
\rangle,
\end{equation}
where the effective potential $V_s(r)=V_{\rm ext}(r) +V_{\rm H}(r)+ V_{\rm xc}(r)$ 
is defined by 
Eq.(\ref{eff_potential}).  We chose a confining potential of
strength $\hbar\Omega=1.6 \ {\rm meV}$ corresponding to that of McEuen {\it et 
al.}\cite{McEuen}.
The Hartree potential for a circularly symmetric quantum dot is given by
\begin{equation}
V_{\rm H}(r)=\frac{e^2}{\epsilon_0 l_B}\int_0^{2\pi}\ d\phi\ \int \ dr'
\frac{r'n(r')}{\sqrt{r^2+r'^2-2rr'\cos\phi + \delta z ^2}},
\label{Hartree_circular}
\end{equation}
where $\delta z$ is a finite thickness  of the 2D electron liquid.
The interaction of the 2D electrons with a gate can also be easily 
incorporated into the Hartree potential.
Although the typical value of $\delta z$ for a real experimental situation is about 
10 nm, in our simulations we took $\delta z \rightarrow 0$ in order to compare our 
results with other theoretical works.  

We have used the LDA 
to obtain the exchange-correlation 
potential $V_{\rm xc}$.  In this approximation, we first need the exchange-correlation 
energy per particle in a homogeneous system $\epsilon_{\rm xc}$ which we chose as
\begin{equation}
\epsilon_{\rm xc}(\nu)=\epsilon_{\rm xc}^{\rm LWM}(\nu)+
\epsilon_{\rm xc}^{\rm C}(\nu).
\label{e: eps}
\end{equation} 
The first term is a smooth interpolation formula of Levesque
{\em et al.,}\cite{Levesque} 
\begin{eqnarray}
\nonumber
\epsilon_{\rm xc}^{\rm LWM}(\nu) & = &
\int^{\infty}_0 dr\ r \left( \frac{e^2}{r\epsilon_0 l_B} \right) [g_{\nu}(r)-1]  \\
& \simeq &
-0.782133\sqrt{\nu}\left(
1-0.211\nu^{0.74}+0.012\nu^{1.7}\right) (e^2/\epsilon_0 l_B)
\label{Levesque_formula}
\end{eqnarray}
for the ground-state energy
obtained by evaluating
the pair correlation functions $g_{\nu}(r)$ at certain fillings $\nu<\frac{1}{2}$ 
for about 256 particles using very 
accurate Monte Carlo methods. 
The second term in Eq.~(\ref{e: eps}),
$\epsilon_{\rm xc}^{\rm C}(\nu)$, contains the cusps in the ground state 
energy which 
cause the FQHE. The  discontinuity in the
slope of $\epsilon_{\rm xc}^{\rm C}(\nu)$ near certain ``magic'' filling
factors $\nu^{\star}=p/q$ is related to the chemical potential gap 
$\Delta \mu=q(|\Delta_p|+|\Delta_h|)$. Here $\Delta_{p,h}$ are the 
quasiparticle (hole) creation energies\cite{Morf_Halperin} at $\nu=\nu^{\star}$. In 
our calculations, we restrict ourselves to include
only the cusps at $\nu=1/3,2/5,3/5$ and $\nu=2/3$, which are the
strongest fractions. (See Appendix A for a detailed description of our
expression for $\epsilon^{\rm C}_{\rm xc}$.) Substituting Eq.~(\ref{e: eps}) into
Eq. (\ref{xc_potential_LDA}) we can find the exchange-correlation potential
as a function of filling factor $V_{\rm xc}(\nu)$ for a 
spin-polarized system.  This potential is depicted in Fig. (\ref{f: exc_corr_spin})
as $V_{{\rm xc},\uparrow}(\nu,\xi=1)$ 
(the generalized spin DFT approach will
be discussed later in Sec. \ref{sudft}). 

The discontinuities in $V_{\rm xc}(\nu)$ in the LDA give rise to a 
numerical instability. The reason is that an arbitrarily small fluctuation in charge 
density close to an FQHE fraction gives rise to a finite change in energy that  
leads to serious convergence problems. To overcome this,
we made the compressibility of the system finite, but very small, 
corresponding to a finite, but very large, curvature instead of a point-like 
cusp in $\epsilon_{\rm xc}$ at the FQHE fractions. This was accomplished by
allowing the discontinuity in chemical potential to occur over an interval in 
filling factor $\delta$ of magnitude $10^{-3}$. This corresponds
to a sound velocity of about $10^6$ m/s in the electron gas, which is
three orders of magnitude larger than the Fermi velocity of a 2D electron
gas at densities typical for the FQHE. In general, the finite compressibility
does not lead to any spurious physical effects so long as the energy of
density fluctuations on a size of the order of the systems size is larger
than any other relevant energy in the problem. The only noticeable
effect is that incompressible plateaus, at which the density would be
perfectly constant were the compressibility zero, have density
fluctuations on a scale of $\delta$. 

We have used our ensemble DFT-LDA scheme to study the edge
reconstruction of a
40-electron quantum dot as a function of magnetic field strength. For a
certain range of magnetic field (typically about 2 -- 3 T),
it is known\cite{Mike_Austr}
that such a dot forms a so-called maximum density
droplet. The maximum density droplet is spin-polarized
with a filling factor which is unity in the bulk and falls off rapidly to
zero at $r_0\approx\sqrt{2N}\ell_B$. It is the most compact droplet
(minimum angular momentum) that can be formed of spin-polarized
electrons in the lowest Landau level.
Increasing the magnetic field increases the importance of 
electron repulsion compared to confinement.
At some higher magnetic field  $B$, the edge of maximum density droplet is 
reconstructed, forming an exchange hole, because it
becomes energetically favorable to
spread out the electron density while still taking
advantage of short-ranged
attractive exchange interaction~\cite{Mike_Austr,Chamon}.

For a slowly varying confining potential,
alternating compressible and incompressible strips in
density may be formed~\cite{Beenakker,Chklovskii} between $\nu=1$ and $\nu=0$ 
regions, with 
the density of the incompressible strips fixed at the density of an FQHE fraction, 
$\nu=p/q$.  The widths of the incompressible strips
are determined by the energy gaps at the fractions $p/q$,
while the widths of the compressible strips
are determined by electrostatics.
These compressible and incompressible strips at the edge of an FQHE system were 
studied in an extended Thomas-Fermi approximation at low but finite 
temperatures by Ferconi, Geller, and Vignale~\cite{Ferconi} for infinite Hall bars.
Their results for the widths of the incompressible and compressible strips
were in good agreement with the predictions by the semiclassical
theories\cite{Beenakker,Chklovskii}. In the extended Thomas-Fermi
approximation,
the kinetic energy was treated as a local
functional, as in the standard Thomas-Fermi approximation, while the 
exchange-correlation energy was included in a LDA. This extended Thomas-Fermi 
approximation is presumably valid in the limit of very slowly varying confining 
potential and large numbers of electrons. 
In contrast, our ensemble DFT approach (Sec.~\ref{ensemble})
treats the kinetic energy
exactly, and does not have any limitations on the number of the electrons.  
Furthermore our calculations can be done at very low or zero temperature.

Figure \ref{f: reconstr} depicts some of our results for a 40-electron
droplet using our ensemble density functional approach. In these
calculations, we chose a parabolic confinement of strength
$\hbar\Omega=1.6$ meV, and $\epsilon_0=12.4$, appropriate for GaAs, and
$T=0$. For values of the magnetic less than about 2.5 T, the
droplet forms a maximum density droplet. At $B\approx2.8$ T, an exchange-hole
forms. This value of magnetic field compares
very well with values obtained from Hartree-Fock calculations. For example,
using $\epsilon_0=12.4$ and $\hbar\Omega=1.6$ meV, we obtain from the work of
MacDonald, Yang, and Johnson\cite{Mike_Austr}
a value of 2.52 T, while the value from the work of de Chamon and
Wen\cite{Chamon} is 2.84 T. As the magnetic field is increased further,
the effective confinement softens, and the droplet undergoes several
reconstructions in which compressible and incompressible strips
are formed. Figure \ref{f: reconstr} shows one example of incompressible
strips at $\nu=2/3$ and $\nu=3/5$ for a magnetic field of 4.1 T. In
our calculations, the incompressible strips form at filling factors
where $V_{\rm xc}$ has a discontinuity. Traversing across an
incompressible strip, the
effective single-particle potential ($V_{\rm ext}+V_{\rm H}+V_{\rm xc}$) is
constant -- as $V_{\rm ext}+V_{\rm H}$ varies, $V_{\rm xc}$ varies across
its discontinuity exactly to screen the change in $V_{\rm ext}+V_{\rm H}$.
Thus, the width of the incompressible strips is determined
by the distance over which $V_{\rm ext}+V_{\rm H}$ varies by an amount
equal to a discontinuity in $V_{\rm xc}$, in agreement with the
argument of Chklovskii, Shklovskii, and Glazman\cite{Chklovskii}.
The widths of the compressible strips are determined by
electrostatics plus the smooth parts of the exchange-correlation potential.

In order to compare our results with exact diagonalization studies of quantum 
Hall dots by Yang, MacDonald and Johnson~\cite{Yang} we have calculated the 
expectation value of the total angular momentum
$\langle M\rangle = \sum_{mn}m f_{mn}$ as a function of the magnetic 
field strength $B$ for $N=6$ spin-polarized electrons in the 
lowest Landau level and a
confining potential of $\hbar\Omega=2.0$ meV.
The results are shown in Fig.
\ref{f: angular_momentum} for two different versions of the
exchange-correlation energy $\varepsilon_{\rm xc}$. The diamonds
($\diamond$) were generated using the Levesque-Weiss-MacDonald
exchange-correlation energy\cite{Levesque}, while the plusses ($+$)
were generated using an exchange-correlation energy due to Fano
and Ortolani\cite{Fano}. This latter is constructed not only
to give the right behavior for $\nu<1/2$, but both the value of
$\varepsilon_{\rm xc}$ at $\nu=1/2$ and particle-hole symmetry
of the lowest Landau level are included to give a good interpolation
formula on the entire interval $0\leq\nu\leq1$ for electrons
in the lowest Landau level. Both exchange-correlation
energies give clear plateaus or plateau-like structures in
angular momentum {\em vs.} magnetic field. However, the
Levesque-Weiss-MacDonald is a rather poor approximation near
$\nu=1$ (a region for which it was not constructed), and
furthermore overestimates the magnitude of the
exchange-correlation potential at about $\nu=1/2$.
As a consequence, the initial maximum density droplet instability
is smeared out and starts at a too low value of magnetic field, and
the formation of a $1/3$ droplet (as is evidenced by studies of the
density profile) occurs at a too high value of magnetic field.
Also, the values of the angular momentum at the plateau-like regions
tend to be too low. For example, the formation of the 1/3-droplet
occurs at $M\approx40$, while the exact value is $M=45$. In contrast,
the results obtained using Fano-Ortolani exchange-correlation energy
tend to be very accurate. For example, the maximum density
droplet instability occurs at $B\approx2.8$ T in our calculations, compared
to $B=2.75$ T in the numerical diagonalizations, and
the 1/3 droplet formation occurs at $B\approx5.3$ T in our calculations,
compared to $B=5.29$ T in the numerical diagonalizations 
(see Fig. \ref{fig:densplot}). In addition,
the plateau-like regions are more developed and flatter in angular
momentum. Still, though, the ensemble DFT tends to underestimate
the angular momentum at the plateaus. We want to
emphasize here that we have not used any adjusting parameters in
our calculations.
Furthermore, the ensemble DFT is not constructed so as to give only integer
angular momentum. Finally, only the energy gaps at
$\nu=1/3$ and $\nu=2/5$, along with their particle-hole
conjugates at $\nu=2/3$ and $\nu=3/5$ were included, while the
numerical diagonalizations used the full Coulomb interaction. Therefore,
perfect agreement between our ensemble DFT results and the numerical
diagonalizations cannot be expected. We also did these calculations at
a finite temperatute of 100 mK ($k_BT/e^2/(\epsilon_0l_B)\sim1\times10^{-3}$), 
which improves the convergence of these small particle systems.
In view of all this, the agreement between our ensemble DFT results and
the numerical diagonalizations of Yang, MacDonald and Johnson\cite{Yang}
must be considered remarkable. We are presently working on extending
the Fano-Ortolani interpolation to include $\nu\geq1$ and several
Landau levels.

\section{Spin textured edge reconstruction of the maximum density droplet.}
\label{sudft}

The edge reconstruction of $\nu=1$ maximum density droplet outlined in 
Sec.~\ref{spdft} suggests that $\nu=1$ is  much simpler to study than the fractional 
quantum Hall filling factors due to the absence of a cusp in
$\epsilon_{\rm xc}(\nu)$ at $\nu=1$.  Therefore, one might expect that $\nu=1$
is a Fermi liquid in the sense that the elementary excitations are well described by 
single-particle excitations and only renormalized by the
interactions.
However,  recent experiments on high mobility GaAs quantum wells~\cite{Barrett} have 
provided evidence for the existence of topological charge-spin textures, 
so-called `skyrmions', near $\nu=1$.  These are non-trivial
many-body excitations due to electron-electron interactions first predicted 
to be the low-energy excitations near $\nu=1$ by Sondhi et al.~\cite{Sondhi}, 
with the energies about half of those of single-particle spin-flip excitations.
The fact that skyrmions are the low-energy excitations near $\nu=1$ (and also
possibly near $\nu=1/3$) raises the possibility of spin-textured edge reconstruction 
of the maximum density droplet.  Therefore, inclusion of the spin
degree of freedom may be essential in the study of inhomogeneous systems.  
Indeed, Hartree-Fock and effective field theoretical calculations have shown
that for a soft confining potential, the edge of an infinite Hall bar at $\nu=1$
becomes unstable to spin-textured reconstruction for weak Zeeman coupling, while 
stronger Zeeman coupling yields a spin-polarized reconstruction~\cite{Karlhede}.

Motivated by these ideas, we have generalized our ensemble DFT approach
to include spin degrees of freedom within the LSDA [Eq.~(\ref{LSDA})].
We have applied this generalization to parabolic  maximum density droplets
to study the edge reconstruction as a function of confinement strength 
(magnetic field) and Zeeman coupling.  We find that as the Zeeman coupling 
is decreased, the edge becomes unstable to spin-textured reconstruction at
a strength of the Zeeman coupling consistent with that found by Karlhede 
{\em et al.}~\cite{Karlhede}
  This provides further evidence for the importance of spin degrees of freedom
in inhomogeneous systems and demonstrates the usefulness of our LSDA ensemble
density functional approach.

For a parabolic dot, the variational principle applied to the KS functional 
[Eq.~(\ref{HKS1})] yields two sets of KS equations for spin up and spin down 
electrons
\begin{equation}
 \left({T}+{V}_{\rm s,\sigma}(r,B)\right)\varphi_{mn,\sigma}(r)=
 \varepsilon_{mn,\sigma}\varphi_{mn,\sigma}(r),
\label{HKspin}
\end{equation}
where
\begin{equation}
   {V}_{\rm s,\sigma}(r,B) = \sigma g^{\star}\mu_0 B
   +V_{\rm ext}(r)+V_{\rm H}(r)+V_{\rm xc,\sigma}(r,B)
\label{ks}
\end{equation}
is an effective potential for the auxiliary non-interacting system.
In the LSDA the exchange-correlation potentials are 
\begin{equation}
V_{\rm xc,\sigma}(r,B)=\left.
\frac{\partial}{\partial n_{\sigma}}
 \left(
 n \epsilon_{\rm xc}[n_{\uparrow},n_{\downarrow},B] 
 \right)
\right|_{n_{\sigma}=n_{\sigma}(r)}.
\label{V_xc_spin1}
\end{equation}
The parametric dependence on the magnetic field $B$ can be incorporated by using
spin filling factors $\nu_{\sigma}=2\pi l_B^2 n_{\sigma}$ as variables instead
of spin densities $n_{\sigma}$. To make connection with the spin-polarized case
we first transform the spin filling factors $\nu_{\sigma}$ to total filling factor 
$\nu$ and spin polarization $\xi$:
\begin{equation}
\begin{array}{l}
         \nu=\nu_{\uparrow}+\nu_{\downarrow}\\
         \xi=(\nu_{\uparrow}-\nu_{\downarrow})/
             (\nu_{\uparrow}+\nu_{\downarrow}).
\end{array}
\label{transform}
\end{equation}   
The exchange-correlation potentials (Eq.(\ref{V_xc_spin1})) then become
\begin{eqnarray}
\nonumber
     V_{{\rm xc},\uparrow} = 
         \frac{\partial}{\partial \nu}
            \left(\nu\epsilon_{\rm xc}\right) +
                      (1-\xi)\frac{\partial}{\partial \xi}
       \epsilon_{\rm xc} \ , \\
     V_{{\rm xc},\downarrow} = 
         \frac{\partial}{\partial \nu}
            \left(\nu\epsilon_{\rm xc}\right) -
                      (1+\xi)\frac{\partial}{\partial \xi}
       \epsilon_{\rm xc} \ ,
\label{V_xc_spin2}
\end{eqnarray}
where the exchange-correlation energy per particle in a homogeneous system with a 
filling factor $\nu$ and polarization $\xi$, {\em i.e.,}  $\epsilon_{\rm 
xc}\equiv\epsilon_{\rm xc}(\nu,\xi)$, has to be approximated. 

We already have given  the expression  for the exchange-correlation energy per 
particle for the spin-polarized electron gas 
$\epsilon_{\rm xc}(\nu,\xi=1)$ [Eq.~(\ref{e: eps})].
The question is then how to obtain a reasonable interpolation formula for 
$\epsilon_{\rm xc}$ between spin-polarized ($\xi=1$) and spin-unpolarized
($\xi=0$) 2D electron liquids for a fixed $\nu$ in a strong magnetic field.
We have constructed a reasonable first approximation, as we now explain.

 In what follows  x(c) as a subscript denotes exchange (correlation) respectively.
We decompose  the exchange-correlation energy $E_{\rm xc}$ into exchange 
$E_{\rm x}$ and 
correlation $E_{\rm c}$ energies.  Since the exchange interaction only acts between 
parallel spins, we  have
\begin{equation} 
E_{\rm x}[\nu_{\uparrow},\nu_{\downarrow}]=\frac{1}{2}
E_{\rm x}[\nu_{\uparrow},\nu_{\uparrow}]+
\frac{1}{2}E_{\rm x}[\nu_{\downarrow},\nu_{\downarrow}].
\end{equation}
Moreover, it follows from dimensional analysis that the exchange energy must scale 
as density (filling factor) to the 3/2 power in a 2D electron gas.  Following  
Oliver and Perdew~\cite{Oliver}, we can then write:
\begin{equation}
E_{\rm x}\sim \int d^2 r \left[
\nu_{\uparrow}^{3/2}(r) + \nu_{\downarrow}^{3/2} (r)\right].
\label{x1}
\end{equation}
We also have from Eqs.~(\ref{transform}) $\nu_{\uparrow}=\frac{1}{2}\nu(1-\xi)$, 
$\nu_{\downarrow}=\frac{1}{2}\nu(1+\xi)$. Equation~(\ref{x1}) 
can then be rewritten as
\begin{equation}
 E_{\rm x}\sim \int d^2 r \ \nu^{3/2}\left[(1+\xi)^{3/2}+(1-\xi)^{3/2})\right].
\label{x2}
\end{equation}
Since in the local density approximation for the exchange energy 
$E_{\rm x}=\int d^2 r \nu \epsilon_{\rm x}(\nu,\xi)$ we are then led to the form
\begin{eqnarray}
\nonumber
\epsilon_{\rm x}(\nu,\xi) & = &\epsilon_{\rm x}(\nu,\xi=1) + 
(\epsilon_{\rm x}(\nu,\xi=0) - \epsilon_{\rm x}(\nu,\xi=1))f(\xi) \\
& \equiv & \epsilon_{\rm x}(\nu,\xi=1) + \Delta \epsilon_{\rm x}(\nu,\xi),
\label{x_form}
\end{eqnarray}
where the function 
\begin{equation}
f(\xi)=\frac{(1+\xi)^{3/2}+(1-\xi)^{3/2}-2\sqrt{2}}{2 -2\sqrt{2}}
\label{interpolation_function}
\end{equation}
is an interpolation function between the two extreme cases $\xi=0$ and $\xi=1$
with $f(0)=1$ and $f(1)=0$.  Although the analogous simple closed
form  for the correlation energy $\epsilon_{\rm c}(\nu,\xi)$ 
is not available, it can be 
always be written as
$\epsilon_{\rm c}(\nu,\xi) = \epsilon_{\rm c}(\nu,\xi=1) + 
\Delta \epsilon_{\rm c}(\nu,\xi)$.
So, as a first approximation we will use  the form  of Eq.~(\ref{x_form})
for the smooth part of the correlation energy $\epsilon_{\rm c}$, too (leaving the
cusps aside for the moment), with the {\em same} interpolation 
function $f(\xi)$, as was suggested first by  von Barth and Hedin~\cite{Barth}. 
Denoting the smooth part of the exchange-correlation energy per particle by
$\epsilon_{\rm xc}^{\rm s}$ we can then write 
\begin{eqnarray}
\nonumber
\epsilon^{\rm s}_{\rm xc}(\nu,\xi) & = & \epsilon^{\rm s}_{\rm xc}(\nu,\xi=1)+
\left[\epsilon^{\rm s}_{\rm xc}(\nu,\xi=0) - 
\epsilon^{\rm s}_{\rm xc}(\nu,\xi=1)\right] f(\xi) \\
& \equiv & \epsilon^{\rm s}_{\rm xc}(\nu,\xi=1)+
\delta\epsilon_{\rm xc}(\nu)f(\xi).
\label{xc_form}
\end{eqnarray}

So far, we have constructed a function $\epsilon_{\rm xc}^{\rm s}(\nu,\xi)$ which
gives a smooth interpolation for the exchange-correlation energy for
any value of $\nu$ and $\xi$. What is left is to add the cusps to this
function. We already have a good approximation for these at $\xi=1$.
We now need to extend this approximation to arbitrary values of $\xi$.
Very little is known about the cusps, {\em i.e.,} the energy gaps, for
arbitrary polarizations. It is known that there {\em is} a gap for 
un-polarized systems at fillings $\nu=2/5$, $\nu=3/5$, and $\nu=2/3$.
The gap, and thus the cusps, occur at very special `magic' configurations
at which the system can take advantage of a particularly low
correlation energy. Therefore, it seems plausible that for a given
value of $\nu$, say $\nu=2/5$, there cannot be an energy gap for any
value of $\xi$ between 0 and 1. In order to incorporate this assumption
into a usable approximation, we interpolate our cusp energy constructed
for polarized systems,
$\epsilon_{\rm xc}^{\rm C}(\nu)$, to arbitrary polarizations by 
multiplying it by a function $g(\xi)$ which is unity at $\xi=0$ and
$\xi=1$ with zero derivative at these points, 
vanishes away from these values of polarization, and is symmetric about $\xi=1/2$. 
All
together, then, we have
\begin{equation}
\epsilon_{\rm xc}(\nu,\xi)=\epsilon_{\rm xc}^{\rm LWM}(\nu)
+\delta\epsilon_{\rm xc}(\nu)f(\xi)+\epsilon_{\rm xc}^{\rm C}(\nu)g(\xi).
\label{eq:exc_spin_2}
\end{equation}
Specifically, we chose
\begin{equation}
g(\xi)=\left[4\xi^2-1\right]^2\left[27-\xi^2\left(40-16\xi^2\right)\right]/27,
\label{eq:g_xi}
\end{equation}
which is the only polynomial in $\xi$ satisfying the above constraints.
Near $\nu=1$ (where there is no cusp in the total exchange-correlation energy),
the sign of the function $\delta\epsilon_{\rm xc}(\nu)$ will then determine the spin 
polarized (ferromagnetic) or spin unpolarized (paramagnetic) ground state of the 
infinite electron liquid (neglecting the Zeeman splitting).  Indeed, substitution of
Eq.~(\ref{eq:exc_spin_2}) into Eq.~(\ref{V_xc_spin2}) gives 
\begin{equation}
\Delta{V_{{\rm xc}}}\equiv V_{{\rm xc},\uparrow} - V_{{\rm xc},\downarrow}=
2\frac{\partial}{\partial \xi}\epsilon_{\rm xc}=
2\delta\epsilon_{\rm xc}(\nu)f'(\xi)+2\epsilon^{\rm C}_{\rm xc}(\nu)g'(\xi).
\label{diff_xc_pot}
\end{equation}
The last term in this expression may be ignored near $\nu=1$. We would
thus expect a 
spin-polarized ground state
if $\delta\epsilon_{\rm xc}(\nu) > 0$ because $f'(\xi) < 0$ 
for all $0\leq\xi\leq 1$ 
so in this case the inequality 
$V_{{\rm xc},\uparrow} < V_{{\rm xc},\downarrow}$ holds. 
Otherwise we would expect a spin unpolarized state.  Since a von Barth-Hedin 
type approximation has been applied before only to the  3D electron gas 
in zero magnetic field, it is natural to ask how faithful this approximation 
is~\cite{Neil_Johnson} to data obtained
from the numerical diagonalization studies~\cite{Zhang,Maksym1} outlined in 
Table~\ref{table1}.  First, reversed spins  are 
in fact possible in the ground state of quantum Hall liquids. It is known from 
numerical studies~\cite{Chakraborty} that the electron gas is 
completely polarized only at filling 
factors $\nu=1,\frac{1}{3},\frac{1}{5}$. Secondly, there are some 
partially polarized ground states that cannot be realized in the approximation used 
here because the  function $f(\xi)$ is monotonic.  If the partially polarized states 
were realized by this function it would have a minimum at some fractional $\xi$.  
However, the simple model Eq.~(\ref{xc_form}) allows us to capture the essential 
physics of the spin unpolarized edge reconstruction of the quantum dot as it will be 
shown on the example below. 

In order to obtain the function $\delta\epsilon_{\rm xc}(\nu)$ in 
Eq.~(\ref{eq:exc_spin_2}) we start by calculating the energy differences 
between spin polarized  and 
unpolarized states for fractions listed in Table~\ref{table1}.
The value for the ground state 
energy of a $\nu=1$ unpolarized system is not available, but a reasonable 
approximation is to take
$\epsilon_{\rm xc}(\nu=1,\xi=0)=
 \epsilon_{\rm xc}(\nu=1/2,\xi=1)=-0.469 \ (e^2/\epsilon_0 l_B),$
implying that the spin-up and spin-down components are uncorrelated.
The ground state energy of a $\nu=1$ polarized system 
$\epsilon_{\rm xc}(\nu=1,\xi=1)= - 0.6265 \ (e^2/\epsilon_0 l_B)$.  
Therefore we have $\delta\epsilon_{\rm xc}(\nu=1)= 0.1575 \ (e^2/\epsilon_0 l_B)$.  
The last quantity is close to the energy required for a exchange-enhanced 
single spin flip, $\sqrt{\pi/8}(e^2/\epsilon_0 l_B)$, at this filling~\cite{Brey}.   
To complete the numerical parameterization of exchange-correlation functional,
we then perform a spline fit to obtain the function $\delta\epsilon_{\rm xc}(\nu)$.  
We have plotted
the exchange-correlation potentials $V_{{\rm xc},\sigma}$ as a function of a 
filling factor $\nu$ at $\xi=1$ and $\xi=0$ in Fig.~\ref{f: exc_corr_spin}. 
We see that at $\xi=1$ the difference between 
exchange-correlation potentials for spin up and spin down electrons $\Delta{V_{{\rm 
xc}}}$ [Eq.~(\ref{diff_xc_pot})] is changing sign 
from negative to positive while the filling factor $\nu$ is decreasing from $\nu=1$ 
to $\nu=2/3$. 
Ignoring the Zeeman splitting and the cusps, 
the ground state of an infinite electron liquid
would change from spin polarized to spin unpolarized. To estimate the possibility of 
having a spin unpolarized state above filling $2/3$ with
the inclusion of the Zeeman splitting, we have to compare the 
dimensionless Zeeman energy $\tilde g$ [Eq.~(\ref{g_def})] with the difference 
$\Delta{V_{{\rm xc}}}$ [Eq.~(\ref{diff_xc_pot})] at this filling 
$\Delta{V_{{\rm xc}}}(\nu=2/3,\xi=1)\approx 0.05 (e^2/\epsilon_0 l_B)$.
 This value is larger then the Zeeman splitting for GaAs $\tilde g\approx 0.02$.
Therefore, the ground state of a  GaAs based homogeneous system is a spin 
unpolarized state at and just above filling factor $2/3$.   
In an inhomogeneous system, in addition to exchange-correlation potential and
Zeeman energy, there are also the Hartree interaction of the 2D electrons and the 
external potential which confines them.  Hence, even in von Barth and Hedin 
type approximation, it is possible to have not only polarized and unpolarized 
states, but also a partially polarized state in inhomogeneous system such a quantum 
dot.

\begin{figure}
\caption{Phase diagram of the edge reconstruction of a parabolic quantum dot in the 
$(\tilde g, \tilde\gamma)$ plane for $N=38$ electrons. Here, 
the confining potential has a strength given by $\hbar\Omega=1.6$ meV.
For $\widetilde\gamma>0.065$,
the system forms a maximum density droplet for all values of
the Zeeman coupling $\widetilde g$. For values of the Zeeman coupling
$\widetilde g$
larger than a critical value $\widetilde g_c$, 
the maximum density droplet undergoes an initial reconstruction to a
spin-polarized exchange hole as the confinement strength $\widetilde\gamma$ is
decreased, while for $\widetilde g<\widetilde g_c$ the maximum density
droplet has a spin-structured instability with decreasing $\widetilde\gamma$.
In these calculations, $\widetilde g_c\approx 0.055$.
}
\label{f: phase_diagram}
\end{figure}
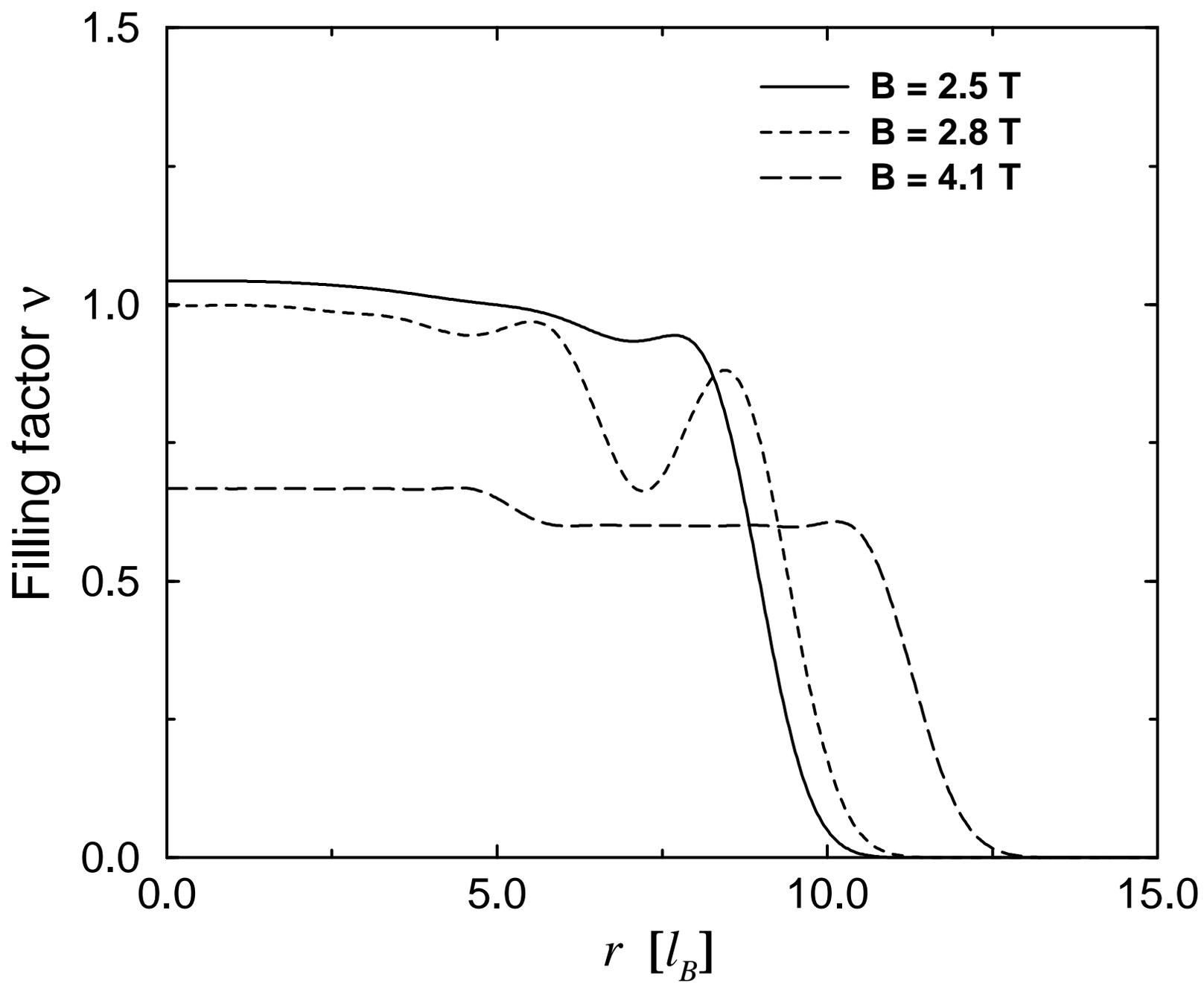
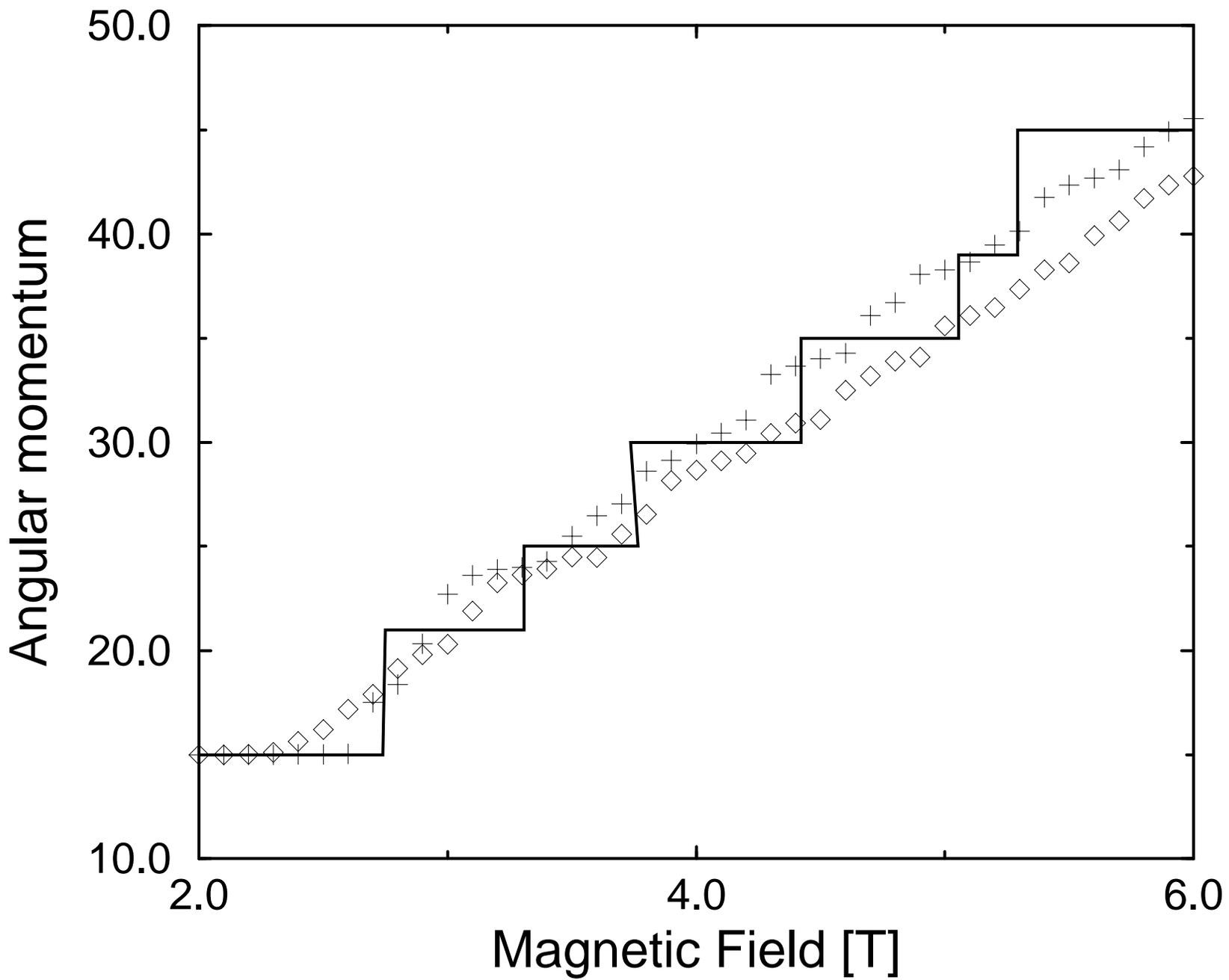
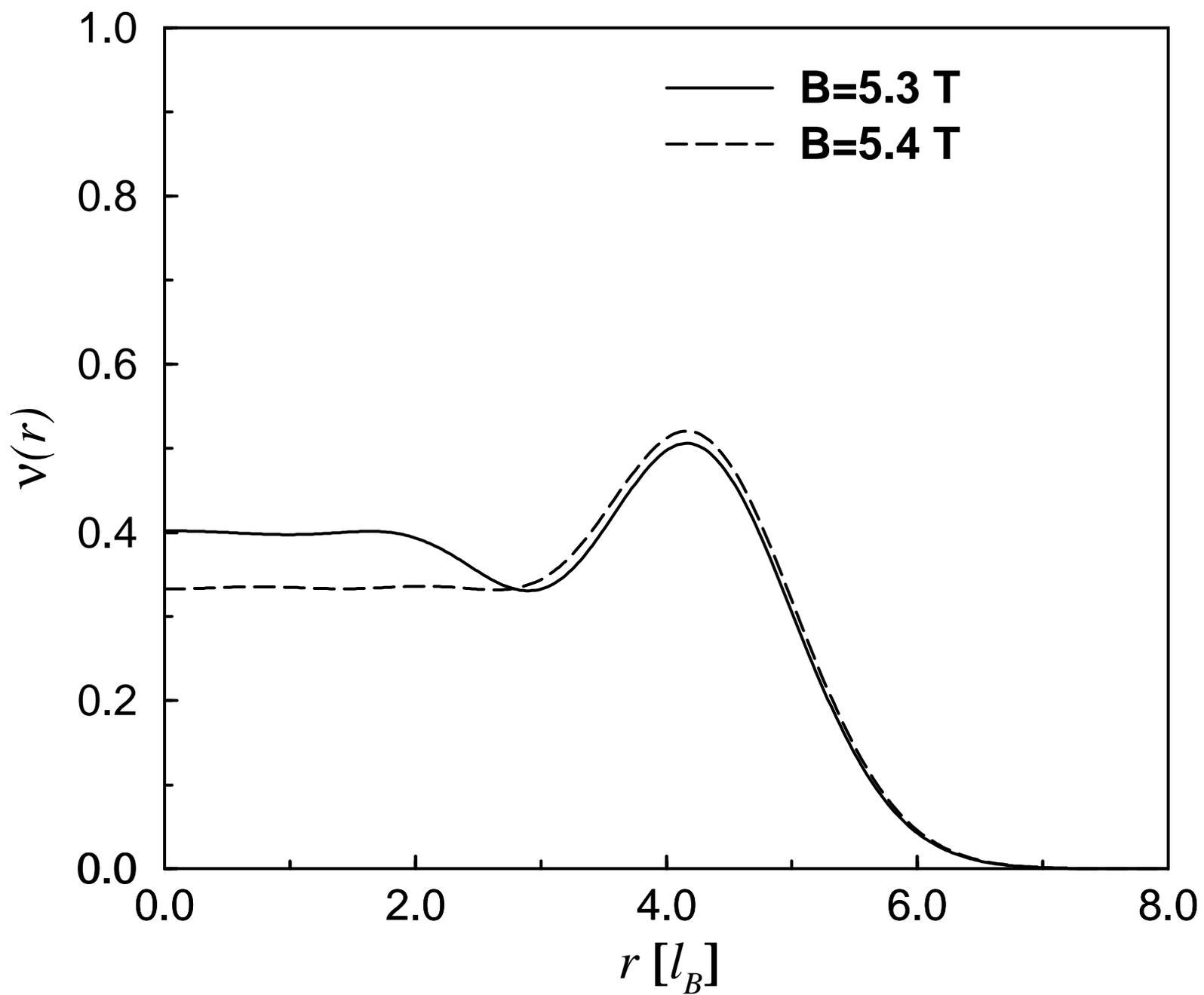
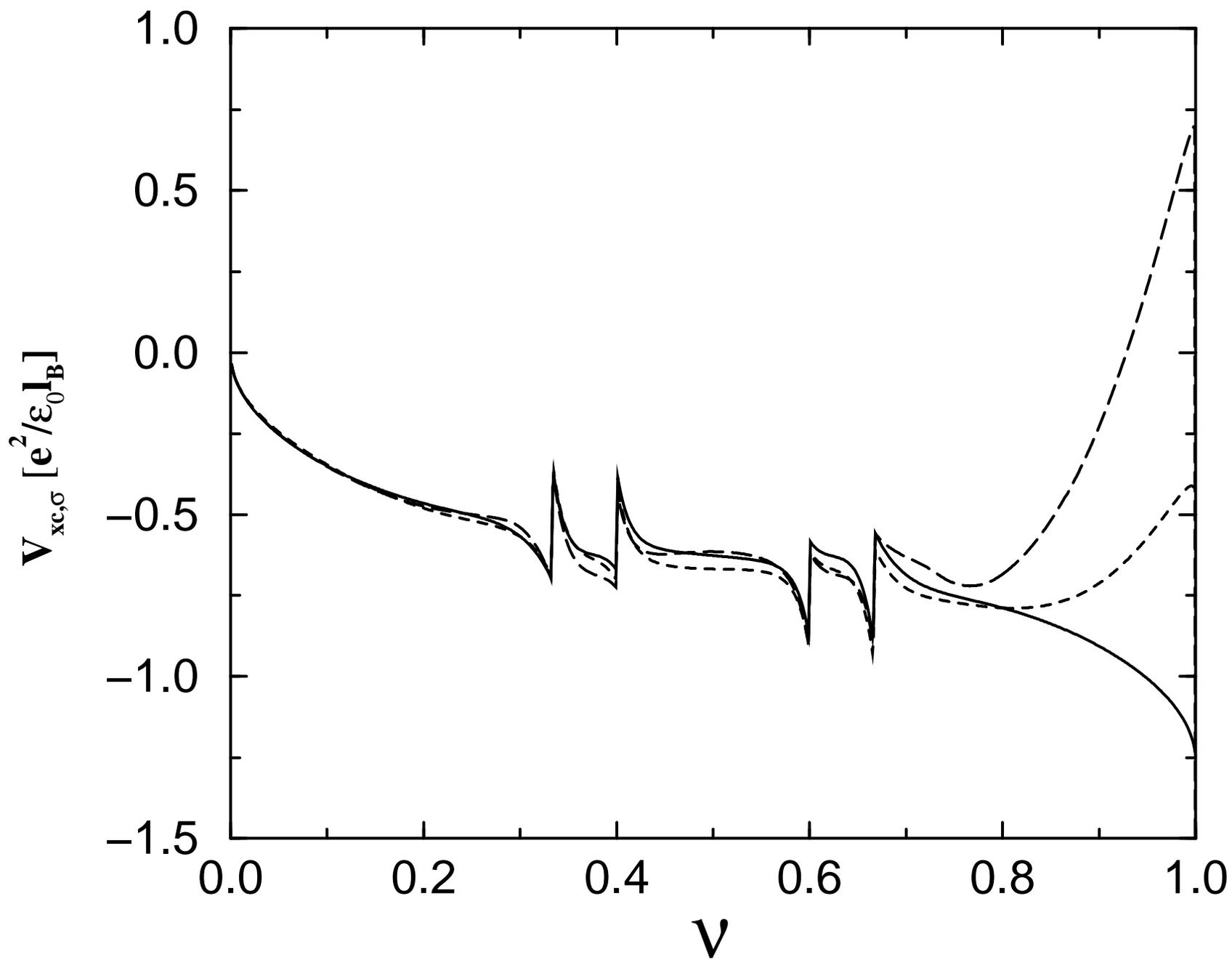
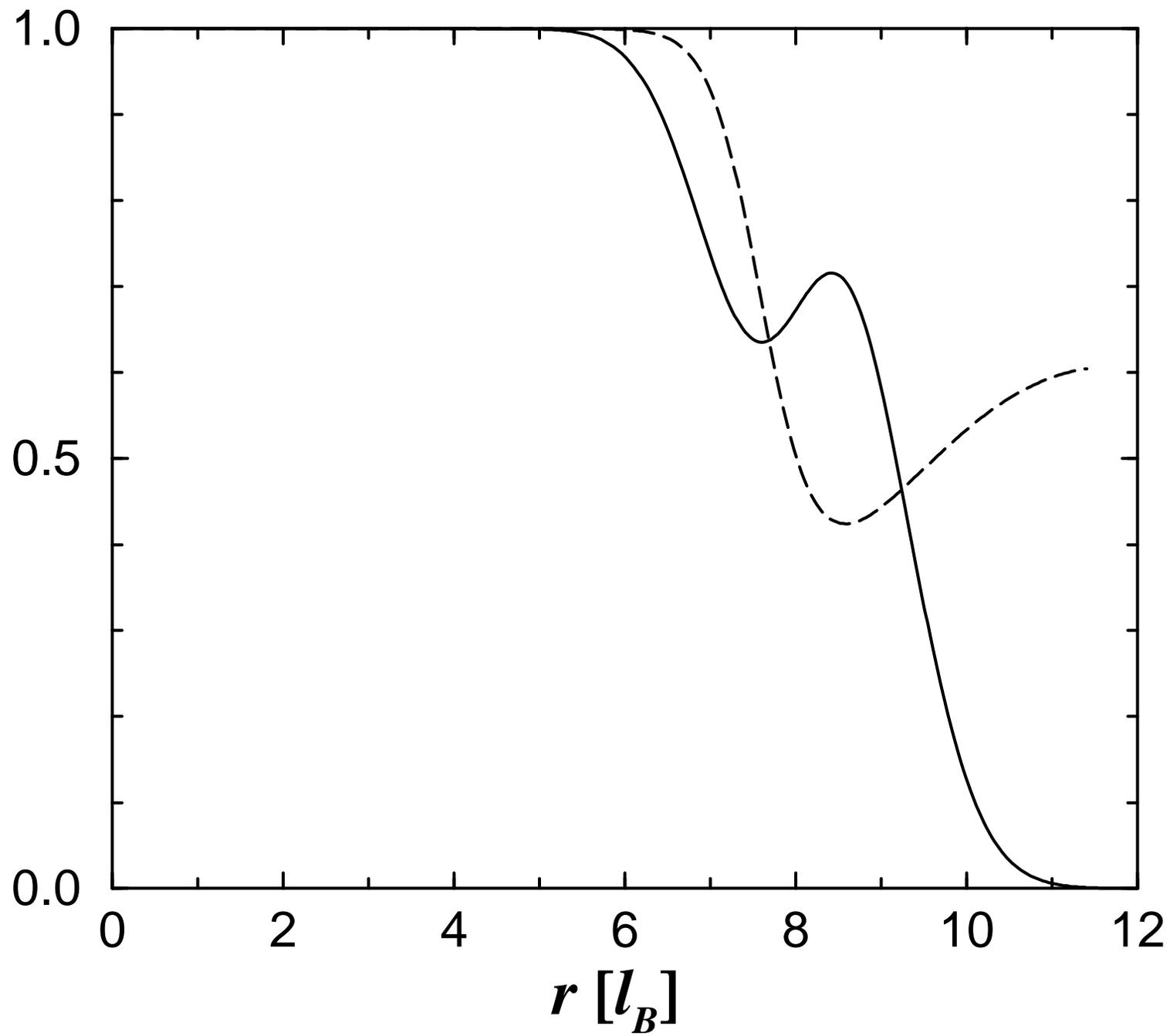
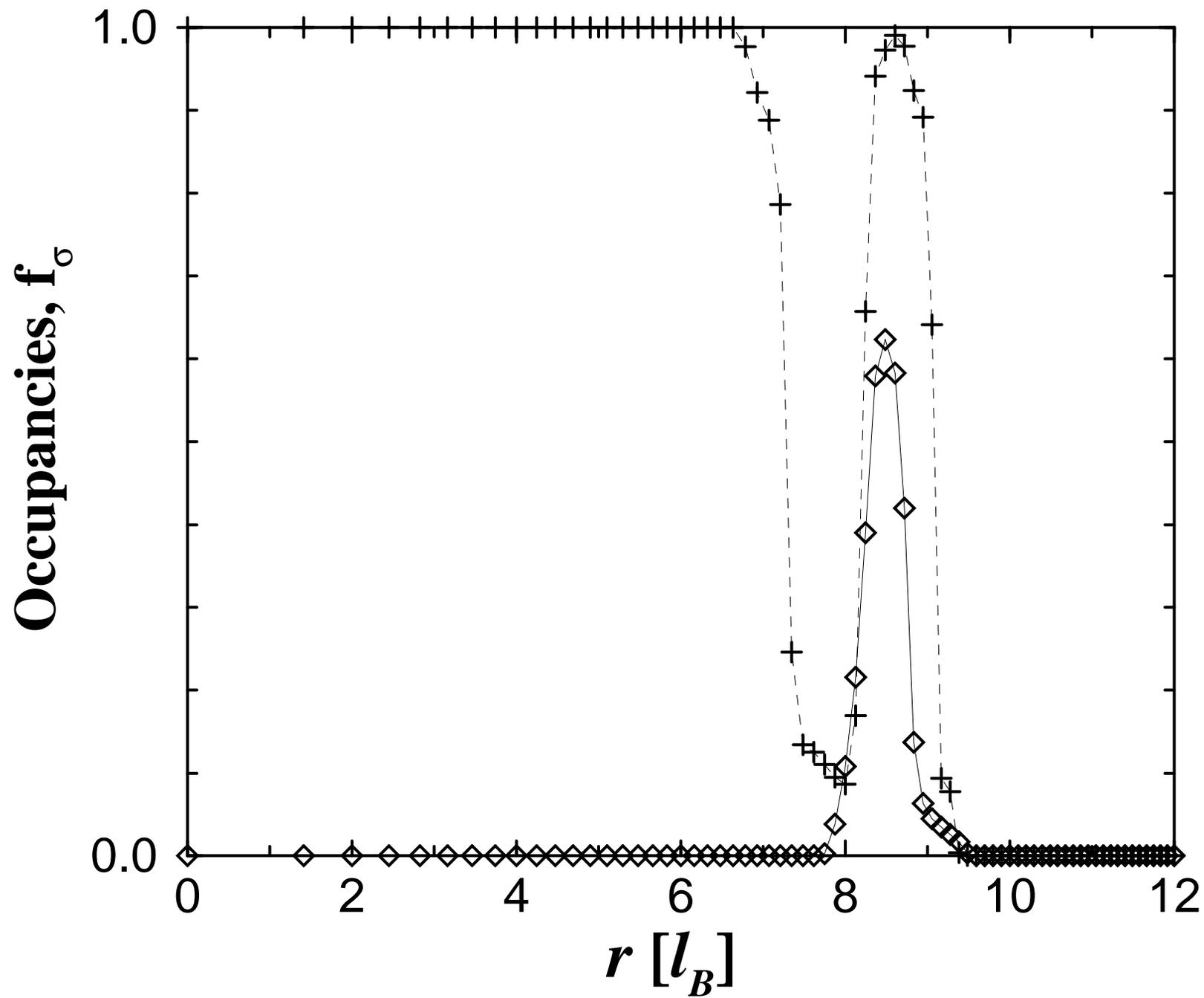
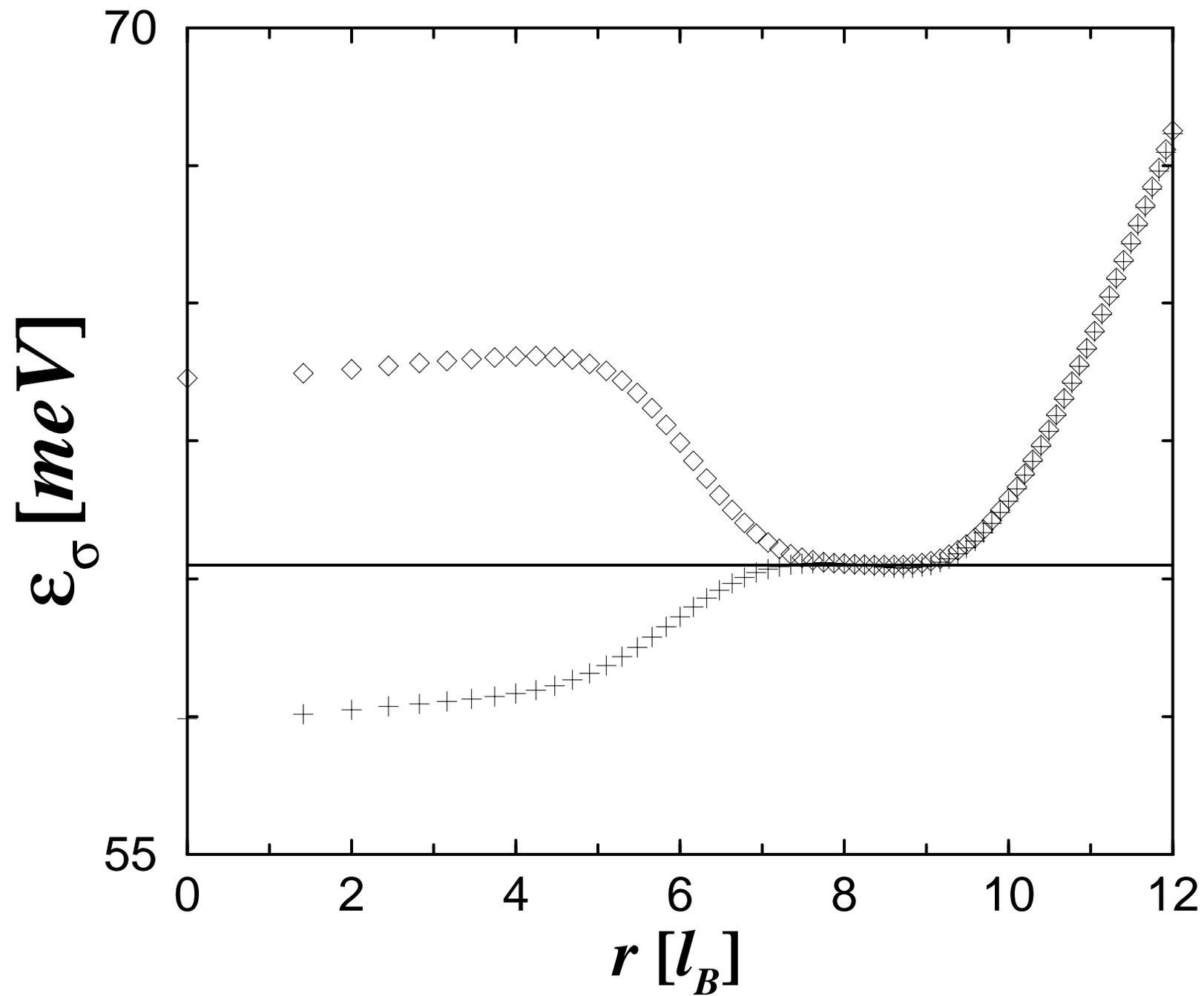
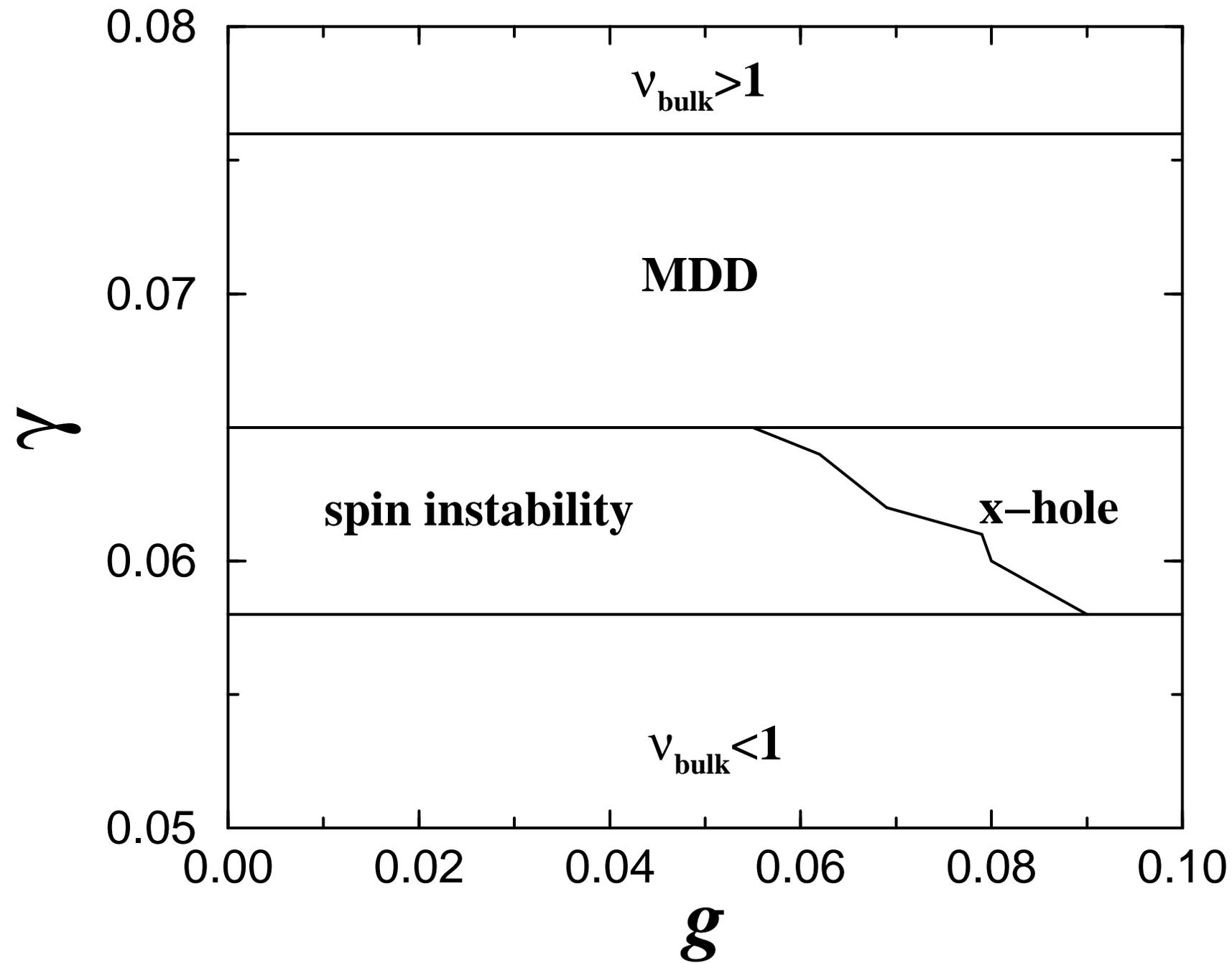

We have investigated the spin density of a quantum dot using our LSDA ensemble 
DFT.  The results, choosing the same parameters
as in Sec.~\ref{spdft}, are shown in Fig.~\ref{f: unpol}.  
 We find that the electron liquid is indeed partially polarized at the quantum dot 
edge provided the Zeeman energy is small enough (Fig.~\ref{f: unpol}). For  Zeeman 
energies below a certain critical value, $\tilde g < \tilde g_c$,  the maximum 
density droplet is reconstructed with the increasing of the magnetic field by 
forming an unpolarized (in general partially polarized) state.
However, as $\tilde g$ increases above the critical value, $\tilde g > \tilde g_c$, 
the polarized reconstruction described in Sec.~\ref{spdft} takes place.  We can plot 
the phase diagram of the edge reconstruction of the quantum dot in the $(\tilde g, 
\tilde\gamma)$ plane (Fig.~\ref{f: phase_diagram}), where 
the dimensionless parameter 
$\tilde\gamma=m^*\Omega^2 l_B^2/(e^2/\epsilon_0 l_B)$ 
characterizes the `softness' of the edge.  
The phase boundaries separate the maximum density droplet, spin polarized and 
unpolarized instabilities.  The value of $\tilde g_c\approx 0.055$ separating
the spin-polarized and spin-structured instabilities is in good agreement
with the value 
$\tilde g^{\rm YMJ}_c\approx 0.03$ found from numerical diagonalization of
parabolic dots by Yang, MacDonald and Johnson~\cite{Yang}.  
A phase diagram analogous to ours was obtained in the work by A. Karlhede {\em et 
al.}~\cite{Karlhede}.  Their value of the critical Zeeman splitting
$\tilde g^{\rm KKLS}_c=0.169$ obtained by Hartree-Fock calculations is about 3 times 
larger then $\tilde g_c\approx 0.055$ from our phase diagram. We speculate that
the difference is due to 
the the fact that correlations were ignored in their Hartree-Fock
calculation, and to the different geometry they used -- an infinite Hall bar.  

\section{Summary and Conclusions}
\label{concl}

We have  developed  a spin ensemble density functional approach to strongly 
correlated systems and used it to study inhomogeneous quantum Hall systems in the
integer and fractional Hall regimes.  For spin-polarized systems, our approach
gives results in excellent agreement with numerical diagonalizations, 
Hartree-Fock and semi-classical calculations.  Note that while all of these latter
approaches have
limited regions of applicability, such as small systems, systems near $\nu=1$, 
or the semi-classical limit, we have here demonstrated that our ensemble density 
functional approach spans all these regions, which makes it a
useful approach to general inhomogeneous quantum Hall systems.

We have generalized the ensemble DFT to include spin degrees of 
freedom within a simple local spin density approximation, and applied this 
generalization to a quantum dot. 
Our results show that for small, but
physical, Zeeman energies, $\tilde{g} < \tilde{g_c}$, 
the maximum density droplet is 
unstable with respect to spin-textured edge reconstructions
as the magnetic field increased. 
At larger Zeeman splittings, $\tilde{g} > \tilde{g_c}$, the maximum density droplet 
is unstable with respect to spin-polarized edge reconstructions. 
Our value of $\tilde{g_c}$ is in good agreement with that obtained from numerical 
diagonalization studies~\cite{Yang}.  
Hartree-Fock calculations for an infinite Hall bar by A. Karlhede {\em et 
al.}~\cite{Karlhede} give a phase diagram qualitatively analogous to ours.
However, Hartree-Fock calculations are limited to $\nu\approx 1$, while our 
ensemble DFT is in principle applicable to general fractional quantum Hall systems, 
e.g. droplets at $\nu=\frac{1}{3}$. 
The accuracy of our approach depends on obtaining good estimates of the 
exchange-correlation energy as a function of both electron density and spin
polarization for homogeneous fractional quantum Hall systems. 
Work is currently in progress to obtain such estimates.  Finally, the spin ensemble
DFT used 
here cannot be used to study spin-charge textures (skyrmions), in which the
spin polarization rotates smoothly in space.  Work is currently under way to
generalize our spin DFT to include such charge-spin textures.

The authors would like to thank M. Ferconi, M. Geller and G. Vignale for helpful
discussions and for sharing their results prior to publication, 
K. Burke and E.K.U Gross for useful comments about the DFT, M. Levy
for a discussion about Ref. \onlinecite{Levy}, and J.M. Kinaret for help
with the LSDA. O.H. would like to 
thank Chalmers Institute of Technology, where part of the work was done.
This work was supported by
the NSF through grants DMR93-01433 and DMR96-32141.

\section*{Appendix}
\label{appendixA}

We will construct the cusp-part of the exchange-correlation energy, 
$\epsilon^{\rm C}_{\rm xc}(\nu)$ for a spin-polarized system by first
considering $\nu<1/2$, and then use electron-hole symmetry to obtain
the form for $1/2<\nu<1$. Finally, for $\nu>1$ we assume 
$\epsilon^{\rm C}_{\rm xc}(\nu)=\epsilon^{\rm C}_{\rm xc}(1-\nu)$.

For spin-polarized systems in the lowest Landau level, we write
$\epsilon_{\rm xc}(\nu)=\epsilon'_{\rm xc}(\nu)+\epsilon^{\rm C}_{\rm xc}(\nu)$,
where $\epsilon'_{\rm xc}(\nu)$ is given by a smooth interpolation, such as
the Levesque-Weiss-MacDonald formula\cite{Levesque}
(although this one does not obey strict
particle-hole symmetry in the  lowest Landau level), or the Fano-Ortolani
formula\cite{Fano}. Particle-hole symmetry yields for the total exchange-correlation
energy
\begin{equation}
\nu\left[\epsilon_{\rm xc}(\nu)-\epsilon_{\rm xc}(1)\right]
= \left[1-\nu\right]\left[\epsilon_{\rm xc}(1-\nu)-\epsilon_{\rm xc}(1)\right],
\label{symmetry}
\end{equation}
from which we obtain
\begin{equation}
\nu\epsilon'_{\rm xc}(\nu)=\nu^*\epsilon'_{\rm xc}(\nu^*)+
(1-2\nu^*)\epsilon_{\rm xc}(1),
\label{symmetry2}
\end{equation}
with $\nu^*\equiv1-\nu$. This means that
\begin{equation}
\nu\epsilon_{\rm xc}^{\rm C}(\nu)=\nu^*\epsilon_{\rm xc}^{\rm C}(\nu^*).
\label{symmetry_cusp}
\end{equation}
We define
\begin{equation}
g(\nu)\equiv\nu\epsilon_{\rm xc}^{\rm C}(\nu).
\label{g_nu}
\end{equation}
Since 
the discontinuities in the chemical potential at fractional QHE
fillings $p/q$ is a relation for $d\left[\nu\epsilon_{\rm xc}(\nu)\right]/d\nu$, it
is easier to work with $g(\nu)$ than with $\epsilon_{\rm xc}^{\rm C}(\nu)$.
Then particle-hole symmetry implies that
\begin{equation}
{d g\over d\nu}=-{d g(\nu^*)\over d\nu^*}.
\label{d_g_d_nu}
\end{equation}
At fractional QHE fillings $p/q$, we must have
\begin{equation}
{d\over d\nu}\left.g(\nu)\right|_{\nu=\left(p/q\right)^+}
-{d\over d\nu}\left.g(\nu)\right|_{\nu=\left(p/q\right)^-}=q(\mu_++\mu_-),
\label{cusps}
\end{equation}
where $\mu_+$ and $\mu_-$ are the quasi-particle and quasi-hole creation
energies (defined to be positive), respectively.
We construct $g(\nu)$ for $0\leq\nu\leq1/2$ to be piece-wise smooth, with
$g(\frac{p}{q})=0$ for $p/q$ a fractional QHE filling, and a discontinuity in
the derivative given by Eq. (\ref{cusps}).

We only included the cusps at $\nu=1/3$, $2/5$, their particle-hole
conjugates, and the corresponding values at fillings inreased by unity.
For $\nu<1/3$, we make the Ansatz
\begin{equation}
g(\nu)=\alpha\nu q\mu_-(p/q)\left(\nu-p/q\right)\left[e^{\left(\nu-p/q\right)}
-g_0\right],
\label{nu_lessthan_.3}
\end{equation}
with $p/q=1/3$ and 
\begin{equation}
\alpha={q\over p(1-g_0)}
\label{alpha}
\end{equation}
and $g_0=\exp(-p/q)$.

For $2/5<\nu\leq1/2$ we take
\begin{equation}
g(\nu)=\frac{5}{a}\mu_+(2/5)\left[1-e^{-a_0(\nu-2/5)}\right],
\label{nu_leq_.5}
\end{equation}
with $a_0=80$.

Next, for $1/3<\nu<2/5$ we used a cubic interpolation
\begin{equation}
g(\nu)=a(\nu-1/3)(\nu-2/5)(\nu-\nu_3).
\label{g_rest}
\end{equation}
Fixing the slope of $g(\nu)$ at $(1/3)^+$ and $(2/5)^-$ then yields
\begin{equation}
a={3\mu_+(1/3)\over (1/3-2/5)(1/3-\nu_3)},
\end{equation}
and
\begin{equation}
\nu_3={5\mu_-(2/5)/3+6\mu_+(1/3)/5\over \mu_+(1/3)+5\mu_-(2/5)}.
\end{equation}

Finally, we smooth out the resulting discontinuities in $V_{\rm xc}$ over
an interval $2\delta$ about the fractional QHE fillings $\nu=p/q$. To do this,
we interpolate linearly $g(\nu)$ between its values at $\nu=p/q\pm\delta$, so
that $d g(\nu)/d\nu=A+B(\nu-p/q+\delta)$, where
$A=g(p/q-\delta)$ and $B=\left[g(p/q+\delta)-g(p/q-\delta)\right]/(2\delta)$.
Simple integration then yields
$g(\nu)=A\nu+\frac{1}{2}B\nu^2-B\nu(\frac{p}{q}-\delta)+C$,
where $C$ is an integration constant given by
\begin{equation}
C=g(p/q-\delta)-A(p/q-\delta)+\frac{1}{2}B(p/q-\delta)^2.
\end{equation}


\newpage
\begin{table}
\begin{center}
\begin{tabular}{|c|c|c|c|c|c|c|c|} \hline
\hspace{2mm} & \multicolumn{5}{|c|}{Potential Energy} &
 $\delta\epsilon_{\rm xc}(\nu)$ &
Ground state \\ \hline
$\nu$  & $\xi=1$  & $\xi=\frac{1}{3}$ & $\xi=0.5$ & $\xi=\frac{2}{3}$ & $\xi=0$ &   
& \\ \hline
$\frac{1}{3}$  & -0.4152 & &-0.4120 & &-0.4135 & 0.0017 & Polarized    \\ \hline
$\frac{2}{7}$  & -0.3870 & &-0.3868 & &-0.3884 &-0.0014 & Unpolarized  \\  \hline
$\frac{2}{5}$  & -0.4403 & &-0.4410 & &-0.4464 & -0.0057& Unpolarized  \\  \hline
$\frac{4}{13}$ & -0.3975 & &-0.3997 & &-0.3970 & 0.0005 & Partially polarized \\  
\hline
$\frac{4}{11}$ & -0.4219 & &-0.4278 & &-0.4241 & -0.0022 & Partially polarized \\  
\hline
$\frac{4}{9}$ & -0.4528  & &-0.4600 & & -0.4554 & -0.0030 & Partially polarized\\ 
\hline
$\frac{2}{3}$  & -0.5232 & -0.5257 & & -0.5291& -0.5331 &-0.0099 & Unpolarized  \\  
\hline
$\frac{3}{5}$  &  -0.5010 & -0.5044 & & -0.5096 &  -0.5074& 0.0064&Partially 
polarized \\  \hline
\end{tabular}
\end{center}
\caption{Potential energy (per particle) for various values of spin
         polarization and $\delta\epsilon_{\rm xc}(\nu)$ (Eq.(~\ref{xc_form}))
         for the four-electron (rows 1-6, Ref.[20]) 
          and six-electron
         (rows 7 and 8, Ref.[46]) 
         systems.  The Zeeman energy is not
         included in the energy values. The unit of energy is
         $(e^2/\epsilon_0 l_B)$.
        }
\label{table1}
\end{table}
\newpage
\begin{figure}
\caption{Edge reconstruction of a spin-polarized quantum dot as the magnetic field 
strength is increased.  Plotted here is the local filling factor 
$\nu(r)$ for a parabolic quantum dot with $\hbar\Omega=1.6$ meV and
40 electrons. For magnetic field strengths $B< 2.5$ T the dot forms a maximum 
density droplet, and for $B\approx2.8$ T, an exchange hole is formed. For stronger 
magnetic fields, incompressible regions form, separated by compressible strips.}
\label{f: reconstr}
\end{figure}
\begin{figure}
\caption{Expectation value of the total angular momentum
$\langle M\rangle = \sum_{mn}m f_{mn}$ as a function of the magnetic 
field strength $B$ indicated by solid line for a spin-polarized 
six-electron droplet
in a parabolic confinement using the Levesque-Weiss-MacDonald (Ref. 37) ($\diamond$) and
Fano-Ortolani (Ref. 42) ($+$) exchange-corrleation energies. The solid shows the 
exact diagonalization studies result from Ref. 10.}
\label{f: angular_momentum}
\end{figure}
\begin{figure}
\caption{Local filling factor as a function of radial coordinate for a six-particle
system in a parabolic external potential with $\hbar\Omega=2.0$ meV. 
Here, the Fano-Ortolani exchange-correlation energy (Ref. 42) was used. 
The transition to a 1/3-droplet occurs between $B=5.3$ T and $B=5.4$ T. The
bumb in electron charge at the edge of the system is characteristic of systems with
a not too soft confining potential.}
\label{fig:densplot}
\end{figure}
\begin{figure}
\caption{The exchange-correlation potentials $V_{{\rm xc},\sigma}$ as a function of a filling 
factor $\nu$ at $\xi=1$ and $\xi=0$ in units of $e^2/(\epsilon_0 l_B)$.  The solid 
line indicates $V_{{\rm xc},\uparrow}$ and short-dashed line corresponds to 
$V_{{\rm xc},\downarrow}$ 
at $\xi=1$.
According to Eqs.~(\ref{diff_xc_pot}) and (\ref{interpolation_function}),  the 
exchange-correlation potentials $V_{{\rm xc},\sigma}$ coincide at $\xi=0$ 
(since $f'(0)=0$) and are shown by the long-dashed line.  
The increase in $V_{\rm xc}$ as functions of $\nu$ at a FQHE filling factors occurs over
a range of a filling factor of 0.002.}
\label{f: exc_corr_spin}
\end{figure}
\begin{figure}
\caption{Spin structured instability at the edge of a quantum dot for a
Zeeman splitting  $\widetilde g=0.014$, magnetic field $B=3.05$ T, and  $N=38$. The 
external potential is
characterized by $\hbar\Omega = 1.6$ meV (so that the dimensionless strength of
the confinement is $\widetilde\gamma=0.063$). 
(a) The solid line depicts the total local filling factor $\nu(r)$ as
a function of radial
coordinate $r$, and the dashed line depicts the polarization.
(b) The occupancies
of the KS states $\psi_{m0,\sigma}(r)$ are plotted against orbital
center coordinate $r_m=(2m)^{1/2} l_B $
 with "+" for majority ($\uparrow$) spin 
occupancies, and  "$\diamond$" for minority ($\downarrow$) spin occupancies. At the
instability of the maximum density droplet for this value of $\widetilde g$, there
is a minority-spin population at the edge of the system.
(c) Eigenvalues of the two 
lowest Landau level KS orbitals, with $+$ depicting eigenvalues of the
majority spin orbitals, and $\diamond$ depicting the eigenvalues of the
minority spin orbitals.   
The chemical potential is indicated by the solid line.  At the edge the filling 
factor takes fractional values, and the KS eigenvalues are here degenerate and
equal to 
the Fermi energy $\varepsilon_F$, in agreement with the general theory of Sec. II.}
\label{f: unpol}
\end{figure}
\end{document}